\documentclass[aps, reprint, superscriptaddress, longbibliography]{revtex4-1}
\usepackage{latexsym,amssymb,amsfonts,mathrsfs,amsmath,bm,ulem}
\usepackage{graphicx}
\usepackage[usenames,dvipsnames]{xcolor}
\usepackage{soul}
\usepackage[linktocpage,colorlinks=true,linkcolor=blue,citecolor=blue,breaklinks=true,urlcolor=blue]{hyperref}

\renewcommand{\vec}[1]{\bm{#1}}

\begin{document}

\title{Evaporative self-assembly of motile droplets} 

\author{Anton Molina}
\affiliation{Department of Materials Science and Engineering, Stanford University, 496 Lomita Mall, Stanford, CA 94305, USA}
\affiliation{Department of Bioengineering, Stanford University, 443 Via Ortega, Stanford, CA 94305, USA}

\author{Shailabh Kumar}
\affiliation{Department of Bioengineering, Stanford University, 443 Via Ortega, Stanford, CA 94305, USA}

\author{Stefan Karpitschka}
\altaffiliation[Current Address: ]{Max Planck Institute for Dynamics and Self-Organization (MPIDS),Am Fa{\ss}berg 17, 3077 G{\"o}ttingen, Germany}
\affiliation{Department of Bioengineering, Stanford University, 443 Via Ortega, Stanford, CA 94305, USA}

\author{Manu Prakash}
\email[Correspondence: ]{manup@stanford.edu}
\affiliation{Department of Bioengineering, Stanford University, 443 Via Ortega, Stanford, CA 94305, USA}

\date{\today}

\begin{abstract}
Self-assembly is the underlying building principle of biological systems and represents a promising approach for the future of manufacturing, but the yields are often limited by undesirable metastable states. Meanwhile, annealing methods have long been an important means to guide complex systems towards optimal states. Despite their importance, there have been few attempts to experimentally visualize the microscopic dynamics that occur during annealing. Here, we present an experimental system that enables the study of interacting many-body dynamics by exploiting the physics of multi-droplet evaporation on a prescribed lattice network. Ensembles of motile binary droplets are seeded into a hexagonal lattice template where interactions are mediated through the vapor phase and can be manipulated through the application of a global gravitational field. We show that for finite systems (61 droplets) the interacting droplets have an effective long-ranged interaction that results in the formation of frustrated, metastable states. Application of a periodic, global gravitational field can drive the system through a non-equilibrium phase transition separating phase-locked synchronization from interaction-dominated behavior. Finally, we directly visualize field-driven annealing that leads to terminal states that are less frustrated. Overall, our results represent a new platform for studying many-body physics with long-ranged interactions, enabling the design of field-based control strategies for programming the self-assembly of complex many-body systems.
\end{abstract}

\maketitle

\section*{Introduction}

Droplets represent an inherently discrete arrangement of matter that is encountered in everyday life and has inspired new ways of thinking about complex systems \cite{shaw1984}. Despite the apparent simplicity of a droplet, a range of complex phenomena often lay hidden in plain sight \cite{cira2015}. In particular, the interactions between multiple evaporating droplets sharing a common vapor phase can give rise to emergent behavior such as spatiotemporal patterns of evaporation and extended evaporation lifetimes that eludes a complete description\cite{wray2019, schafle1999, lacasta1998, carrier2016, khilifi2019, pandey2020}. Part of this complexity emerges from the long-ranged nature of the vapor concentration surrounding an evaporating droplet, which decays with a $1/r$ scaling \cite{giansanti2008, mukamel2008}. Thus, multiple droplets sharing a common vapor phase can effectively "sense" each other over long distances. However, droplets can also act as both sources and sinks of vapor\cite{carrier2016, pandey2020}. Therefore, the long-ranged interactions are intrinsically non-additive and can give rise to non-trivial behavior observed in more familiar multi-body systems like protein-folding and frustrated magnets \cite{dill1997, ouyang2016, giansanti2008, islam2013}. Geometry can further couple into the system via deformations. For example, changes to droplet shape can lead to geometry dependent evaporation profiles independent of many-body effects \cite{saenz2017, pandey2020}. When droplets are composed of two different miscible liquids, evaporation can be coupled to motility. Therefore, evaporation represents a currently unexplored driving force for self-organization \cite{cira2015, benusiglio2017}. Meanwhile, gravity can be used as an external field to manipulate droplet motion\cite{benusiglio2018}. The combination of many interacting elements with competing forces is expected to produce a rich variety of complex behavior that is amenable to direct observation \cite{grzybowski2004}.

The macroscopic nature of droplets make them ideal candidates for building experimental tabletop systems since they allow for direct visualization of otherwise inaccessible phenomena. For example, annealing methods are conceptually very important for a variety of fields yet there have been few experimental attempts to visualize the underlying dynamics\cite{kirkpatrick1983, morgenstern1987, salamon1988, glauber1963}. Annealing is a term with origins in metallurgy that has been used to describe a process of controlled cooling to reduce the number of defects during crystallization by making progressively fewer excursions to higher energy states. High energy excursions can be realized either through thermal fluctuations or by applying a global field. Kirkpatrick \textit{et al.} recognized that this process represented a highly general algorithm that could be applied to many combinatorial optimization problems\cite{kirkpatrick1983}. Meanwhile, field-based annealing has been used to produce high quality colloidal crystals and find the ground state of artificial Ising systems \cite{aksay1996, wang2006, schiffer2019}, inspiring new approaches in combinatorial optimization \cite{pal2002}. While artificial Ising spins have provided many insights into understanding the dynamics of thermal relaxation processes\cite{farhan2013NatPhys, farhan2013PRL}, very little has been reported on the dynamics during annealing, either thermal or field-induced. Studies on monolayers of colloidal particles have provided confirmation and direct visualization of the Kibble-Zurek mechanism during thermal annealing\cite{keim2015}. However, the usefulness of such an insight to finite-systems with geometric constraints or if similar results hold for field-induced annealing is unclear. Meanwhile, an increasing ability to observe and manipulate matter creates a need to develop more sophisticated control strategies that integrate information beyond bulk quantities\cite{sood2020, tang2016, aliprandi2015, fukui2016}.

Here, we present a table top many-body lattice system with long-ranged interactions using motile, binary droplets as the fundamental interacting unit \cite{cira2015, benusiglio2017, benusiglio2018, karpitschka2017}. Binary droplets are a class of synthetic active matter that is formed from two well-chosen miscible liquids that exhibit evaporation induced surface tension gradients leading to an apparent contact angle due to Marangoni contraction \cite{cira2015, karpitschka2017}. When a droplet is placed in an asymmetric vapor field, radial symmetry in surface tension is broken resulting in a droplet motility subject to negligible surface pinning. Binary droplets are therefore capable of translating a time-evolving vapor field into complex dynamics. Binary droplets can be made from low-cost, non-toxic, and easily accessible materials like water and propylene glycol, making them ideal for developing a model system for exploring long-ranged interactions. The physical properties of binary droplets allow us to use patterns of hydrophobic material to confine individual droplets to a well defined unit cell, eliminating coalescence. This allows us to treat each droplet as a particle confined to a potential energy well but free to interact with neighboring particles through a vapor mediated potential. We find that the interacting droplets have an effective, long-ranged interaction that results in the formation of frustrated, metastable states. Application of a global, gravitational field with a time-dependent orientation can drive the system through a non-equilibrium phase transition separating phase-locked synchronization from a frozen regime dominated by droplet-droplet interactions. Finally, we show that field-induced annealing can lead to terminal states that are less frustrated and lower in energy than systems prepared without the application of an annealing schedule. This experimental platform allows for the direct visualization of the microscopic degrees of freedom of a strongly interacting many-body system subject to external fields.

\section*{Experimental Description}
In our experiments, we create a system of interacting droplets on a hexagonal honeycomb lattice with edge width $w=1$ mm and lattice constant $a=7.5$ mm (\textbf{Figure 1a}). The binary droplets are composed of water and propylene glycol and have average radius of $R=2.3$ mm with standard deviation $\sigma=0.2$ mm. Experiments are carried out under UV illumination and droplets are visualized by the addition of a small amount of UV-active dye ($0.08 \%$) $[V/V]$. Data is captured using a digital camera with a long pass filter, giving the droplets a green color. In a hexagonal honeycomb geometry, a droplet can be described as being in one of six states corresponding to the six vertices of a hexagon. Alternatively, these vertices correspond to the lattice sites of the dual lattice and are lower in energy than non-vertex sites. However, droplet-droplet interactions limit the number of accessible states. For example, the equilibrium structure of the smallest possible lattice ($N=3$) is a cluster of three droplets (triplet) about the common vertex since this is where the concentration of vapor is highest and a droplet experiences a net force of order $\sim 10$ $mN$ (\textbf{Figure 1b}). In the context of geometrically frustrated systems, this is the locally preferred structure. Frustration emerges as the lattice size is increased and the locally preferred structure can no longer be propagated globally, resulting in the formation of defect sites. In the $N=7$ lattice, we expect to observe only a single triplet and two vertices occupied by doublet structures. This scenario resembles the classic picture of frustration in antiferromagnetic spin systems on a triangular lattice. Here, we study the $N=61$ lattice which has a non-trivial state space with $6^{61}$ or $\sim 10^{47}$ unique states; however, droplet-droplet interactions constrain this space, making only a subset of these states physically accessible.

We begin by investigating the self-assembly of the system initialized in a high-energy state (\textbf{Figure 1c-e; Supplementary Video 1}). A high-energy state is realized by depositing 61 droplets in parallel at the center of each unit cell (\textbf{Figure 1c}). In this configuration, droplet-droplet distances are maximized corresponding to a maximum in the potential energy. The assembly process begins several seconds after the droplets are deposited on the surface. During this time droplets form an effective contact angle by means of an evaporation-induced surface tension gradient. We perform the experiment in an enclosed chamber, minimizing any disturbances in the vapor field due to ambient air currents and enabling a steady state to be reached (\textbf{SI Figure 1}). This allows us to approximate the propagation of the vapor based potential as a diffusion dominated process. A lattice with size $\sim$ $10$ $cm$ would require $\sim$ $200$ $s$ to establish a steady state. This timescale sets a lower bound on our experimental observations. We observe the system for 20 minutes, corresponding to a change in the average droplet diameter of $\sim 5 $ \% \textbf{SI Figure 2}. The absence of disturbances in the vapor field is equivalent to the absence of thermal noise; therefore, this self-assembly process is analogous to a rapid-quench where we can expect a suboptimal assembly product with numerous frozen in defects (\textbf{Figure 1d}). While most of the assembly occurs within 5 minutes, longer observation times reveal rearrangements, suggesting the presence of multiple timescale relevant to the assembly process (\textbf{Figure 1e}).

In order to better understand this system, we develop a simple, numerical model that captures the key features of this system. Previous work has described the vapor concentration field as a phase separation process governed by Cahn-Hilliard dynamics\cite{lacasta1998, schafle1999}. This description accounts for the time evolution of the field and can describe the cooperative transport of vapor within the system. Here, we take a simpler approach and assume that the system is in steady state. It has been established that droplets separated by a center-to-center distance $r$ experience a $1/r^{2}$ attractive force that is caused by asymmetries in the surrounding vapor concentration field $\phi$ \cite{cira2015}. This allows us to identify the vapor field established by a droplet in the system as contributing to a potential field. Here, we model the interactions using a two-body potential where the $i^{th}$ droplet in the lattice with radius $R$ can be modelled as a source of vapor with field\cite{eggers2010, carrier2016}:

\begin{equation}
    \phi(r_{ij}) = 
    \begin{cases}
    1 & r_{ij} < R \\
    \frac{2}{\pi} \arcsin{(\frac{R}{r_{ij}})} & r_{ij} \geq R
    \end{cases}
\end{equation}

\noindent
By defining $\mathcal{N(\xi)}$ to be the number of interacting neighbor droplets in an interaction shell that account for all droplets $\xi$ discrete lattice sites away, we obtain an expression for the system energy $E$ that accounts for non-additivity due to vapor screening by varying $\xi$:

\begin{equation}
    E = \sum_{i=1}^{N} \sum_{j = 1}^{\mathcal{N (\xi)}} \phi(r_{ij}) 
\end{equation}

\noindent
This is a steady-state, energy based model that makes many assumptions. First, we ignore the diffusion timescale and assume that the vapor field is established instantaneously. We also assume the droplets to be hard disks of uniform size with zero contact line pinning. Finally, these experiments occur in a low Reynolds number limit we can neglect inertia and obtain the equations of motion directly from the over damped gradient equation $ \dot{\vec{r}} = - \zeta \nabla E$, where the value of the damping parameter is $\zeta = 1$. To obtain a numerical solution of this stiff ODE system, we utilize a backward difference formula. \textbf{Figure 1f} shows the numerical representation of this vapor potential. A triangular lattice is overlaid showing that the dual lattice sites correspond to sites where the vapor field is locally maximal.

\section*{Results}

\subsection*{Rapid Quench}

We find that there are four stages of the self-assembly process corresponding to two time scales (\textbf{Figure 2a, Supplementary Video 2}). In the initialization stage, droplets wet the surface on the seconds timescale and form an effective contact angle during the first $  t \sim 30 s $. Once an effective contact angle is established, motility is observed and the droplets begin navigating the complex vapor landscape produced by the neighboring droplets. An isolated pair of droplets will take $\sim 20 s$ to travel the length of one lattice constant \cite{cira2015}. In stage I ($ t \sim 120 s $), we observe an inward collapse resembling the gravitational Jeans instability expected for systems with long-range attractive interactions \cite{ramaswamy2014, golestanian2012}. Droplets rush towards the center of the system and are stopped by the hydrophobic boundary but not necessarily at a vertex site. Stage II ($ t \sim 300 s $), coinciding with the time required for the diffusion of water vapor to span the system, nearly all droplets settle into well defined vertex states. Importantly, the hexagonal template serves to deflect the interacting droplets in the direction of maximum vapor concentration that is consistent with the template geometry. We note that some droplets take much longer to arrive at a vertex state than others. Since many of these slower droplets occur in the bulk of the system, their velocity may be attributed to the presence of shallow vapor gradients due to conflicting long-ranged interactions. Stage III ($ t > 300s $) is characterized by rearrangements of the system from one well defined vertex state to another. This can be understood in terms of local residual gradients giving way to the gradient established by the final arrangement of droplets. Alternatively, they could also reflect changes in the vapor field due to cooperative evaporation effects whereby the spatial distribution of vapor changes as a complex function of time. 

Examining the time evolution of the vertex statistics, we see that stage I is characterized by a rapid increase in singlets and quickly saturates (\textbf{Figure 2b}). This high population of energetically unfavorable singlets reflects the idea that this assembly process is very similar to a rapid quench. Additionally, we also observe a rapid increase in doublets, indicating that they form nearly instantaneously. Stage II coincides with the formation of triplets through the combination of singlets and doublets; triplet formation from three singlets simultaneously was never observed in experiment. In stage III, we see rearrangements in doublet and triplet populations while the singlet population remains relatively stable. All of these rearrangements correspond to a monotonically decreasing system energy, consistent with a zero-fluctuation system.

While a small amount of geometric frustration is to be expected in this system due to finite-size effects, we observe a degree of frustration that is far in excess of what can be understood from geometry alone. Given the long-ranged nature of the attractive potential, we expect that the frustration due to competing long-ranged interactions to be substantial \cite{islam2013}. However, since the evaporation of multi-droplet systems is a complex, cooperative \cite{schafle1999, carrier2016, lacasta1998, pandey2020} process where droplets act as both sources and sinks of vapor, we cannot \textit{a priori} predict the actual value of $\xi$ in Equation 2. We therefore take $\xi$ as a discrete fitting parameter to generate statistical data that can be compared against experiment. Frustration is quantified using a misfit parameter $\mu$ introduced by Kobe et al. \cite{kobe1995}. The energy $E_{i}$ of an observed, final configuration is compared with the energy of idealized configurations where all interactions are satisfied $E^{id}_{min}$ and where all interactions are unsatisfied $E^{id}_{max}$:

\begin{equation}
    \mu(E_{i}) = \frac{E_{i} - E^{id}_{min}}{E^{id}_{max} - E^{id}_{min}}
\end{equation}

\noindent

Thus frustration can be quantified by a scalar value bounded between 0 and 1. We numerically explore the effect of changing $\xi$ up to 5 neighbors and confirm that increasing the range of interactions results in higher levels of frustration (\textbf{Figure 2c}). Considering only nearest-neighbor interactions gives the highest average number of triplets; triplet formation even occurs at the boundary. Increasing $\xi$ increases the level of frustration, suppressing the formation of vertex structures and increasing the extent of edge effects. Performing this experiment many times ($n=33$), allows us to calculate average values for each of the different vertex structures, giving $<n_{triplets}> = 3.2$, $<n_{doublets}>= 5.7$, $<n_{singlets}> = 39.5$ (\textbf{Figure 2d}). These statistics yield a misfit value of $\mu = 0.69$ corresponding to a value of $\xi=3.5$, motivating our choice of $\xi=4$ to simulate the system. This suggests that droplets may act to screen one another up to a certain distance, supporting the observation that long interaction distances corresponds to a level of frustration not observed in experiment.

\subsection*{Manipulation of energy landscape with global field}

We next explore the ability to drive droplet dynamics with time-varying global fields. The forces involved in droplet attraction are of such magnitude that gravity can be used to compete with droplet-droplet interactions. A gravitational field exerts a force $F_{g}=\rho V g \sin{\alpha}$ on all droplets simultaneously to an extent determined by the droplet volume $V$ and the angle of tilt $\alpha$. Previous work on pairs of binary droplets has shown that the response of the droplets to an external field is nearly instantaneous due to negligible surface pinning \cite{benusiglio2018}. Therefore, we might expect that droplet volume acts as a source of time-dependent quenched disorder by defining a susceptibility to an external gravitational field.  Furthermore, the lattice geometry imposes an anisotropy at the scale of individual vertices whereby there is a particular orientation $\theta$ of the gravity vector relative to the lattice along which the force required to break a triplet into a doublet and singlet is minimal (\textbf{SI Figure 3a-b}).  Considering an isolated triplet, we can calculate the angle $\alpha_{min}$ required to give a force sufficient to break a triplet structure associated with a hydrophobic barrier of width $w=1$ mm. This angle was measured experimentally to be $\alpha_{min} = 3.5 \deg$ (\textbf{SI Figure 3c-d}), in good agreement with order of magnitude calculations. We use this value to calculate a non-dimensional field strength $\gamma = \alpha / \alpha_{min}$. 

In the following experiments, a rotary actuated Stewart platform is used to impose a time-varying field on the droplet system (\textbf{SI Video 1, SI Figure 4-5}). The magnitude $\gamma$ of the gravity vector is varied from 0.0 to 1.61 and its orientation $\theta(t)$ has a sinusoidal time dependence $\sim sin(\frac{2 \pi t}{T})$ where $T$ is the precession period and fixed at 60 s. This functional form samples all orientations of $\theta$ relative to the lattice equally, in contrast to a thermal field where sampling will be stochastic. We expect there to be two regimes: at high $\gamma$ droplet motion will track the field and at low $\gamma$ the gravitational force will be too weak to overcome droplet-droplet attraction and unable to drive dynamics. 

We define two order parameters to characterize this athermal non-equilibrium transition. We recognize that a strong field imposes an oscillatory frequency, creating a system of coupled oscillators that are linked to one another through the vapor phase providing a non-linear restoring force. The coherence $z$ of coupled oscillators is commonly characterized by the Kuramoto order parameter, averaging over the phase behavior of individual droplets:

\begin{equation}
    z(t) = \frac{1}{N} \sum^N_j  e^{i \phi_j}.
\end{equation}

\noindent
Experiment reveals that the dependence of this order parameter on $\gamma$ follows a smooth second-order like transition (\textbf{Figure 3a, SI Figure 6-7, SI Video 3}) similar to that observed by Viscek \textit{et al.} \cite{viscek2002} who numerically studied a system of coupled oscillators with an explicit spatial dependence. This transition is captured by our numerical model. However, we obtain better agreement by setting $\xi=1$. This result can be understood by noting that diffusion sets a timescale for signal propagation in the system. The importance of this timescale relative to droplet motion can be characterized through a Peclet number ($Pe = \frac{aU}{D}$). A droplet with velocity $U \sim 0.3 \frac{mm}{s} $ and diffusivity $D = 2.42 \frac{mm}{s^2}$ with characteristic length scale set by lattice constant $a$ is characterized by $Pe \sim 1$ indicating that both advective and diffusive processes are important. Put differently, fast-moving droplets display short-range effective dynamics characterized by $\xi=1$. Furthermore, accounting for disorder in droplet size of 10 \%, reduces the level of coherence, providing a better fit to data, suggesting that droplet disorder plays an important role in droplet dynamics. This effect becomes more apparent upon inspection of the microstates of the system, where further insights than can be obtained from average quantities alone (\textbf{Figure 3b, SI Figure 8, SI Video 4}). At high $\gamma$, we see coherent phase behavior in the bulk of the system. Additionally, droplet trajectories tend to be more circular towards the center reflecting the radial symmetry of the global vapor field. Phase defects are observed at the boundary and at a few sites in the bulk. Phase defects always begin at the boundary since these sites have fewer neighbors. Meanwhile, numerical simulations show that phase defects in the bulk require the presence of disorder since smaller droplets are less susceptible to driving. These results are similar to studies on other athermal non-equilibrium systems subject to global driving \cite{budrikis2011, budrikis2012}.

The transition between field- and interaction-driven regimes can also be characterized in terms of the period-averaged distance between nearest-neighbor droplets $u$. A strong field will eliminate any deviations from a uniform bond length that might arise due to droplet-droplet interactions. Furthermore, we expect that the system average $ \langle u \rangle $ will have a weak dependence on $ \gamma $ since the system is geometrically constrained and any increase in one bond length must coincide with a reduction in another. This implies that an increasing standard deviation in $ \langle u \rangle $ is diagnostic of symmetry breaking due to droplet-droplet interactions. Furthermore, having a distribution of droplet radii will increase this average value and broaden the distribution in $u$. In \textbf{Figure 3c}, we see that experiment compares well with numerical simulations for $\xi = 1$. We observe that $ \langle u \rangle $ is systematically greater than the numerical prediction in the strong-field regime, reflecting deformations to the droplets as their center of mass is compressed into a vertex site. We also observe the expected transition in the standard deviation in $\langle u \rangle$. We can understand this transition in terms of the observable microstates, which allows us to directly visualize the spatial distribution of bond lengths (\textbf{Figure 3d, SI Figure 8, SI Video 5}). At high $\gamma$, the system has a narrow distribution of bond lengths with heterogeneity coming from an elastic mode at the boundary. As the field is reduced, heterogeneity proceeds into the bulk, broadening the distribution. Interaction-induced elasticity prevents the distribution from splitting into a bimodal distribution. At low fields, the standard deviation increases dramatically, corresponding to the formation of fixed vertex structures. Here, the numerical results for $\xi = 4$ gives better agreement confirming the importance of long-range interactions at low-field strengths. 

\subsection*{Field-induced annealing}

The ability to drive a system through a transition using a single control parameter allows for implementation of an annealing protocol. The exact temporal sequence $\gamma(t)$ defines the annealing schedule and its time dependence in thermal systems is an important theoretical problem. Here, we approximate the heuristic described by Morgenstern \textit{et al.} and begin with the system in a phase-locked state and reduce the field to values of $\gamma$ where competing droplet-droplet interactions lead to non-trivial dynamics. We remain at each value for 3 periods which is sufficient to reach quasi steady state. 

As before, we observe that vertex structure formation begins at the boundary and follows the field. However, as the field is lowered, phase coherence is reduced and vertex structures are no longer able to follow the field. In the representative example shown in \textbf{Figure 4a}, we see that the motion of vertex structures is limited to one side of the system. Interestingly this side coincides with a high number of large droplets (\textbf{SI Figure 10}). Further reducing the field, we see the formation of doublet structures that are converted into triplets as the field strength is further reduced. Following the energy of the system over time shows that the system tends towards lower energies as $\gamma$ is reduced but makes excursions to higher energy states at a frequency set by the global field (\textbf{Figure 4b, SI Video 6}).

We can gain further insight into the annealing process by considering correlations and fluctuations in the system. An equal time correlation function is defined:

\begin{equation}
    C(\xi) = \frac{1}{T} \int_0^T 
    \langle
    \cos(\theta(\vec{r}, t) - \theta(\vec{r} + \xi, t)
    \rangle_{\xi}
    dt
\end{equation}

\noindent
where $\langle . \rangle_{\xi}$ represents an expectation value over interactions that are $\xi$ lattice sites away for different possible values of the position vector $\vec{r}$. The correlation function can be fit with an exponential $\sim e^{\frac{\xi}{l}}$ where $l$ corresponds to the correlation length. As expected, the correlation length diverges at high $\gamma$ (\textbf{Figure 4c}). At intermediate $\gamma$, we obtain a correlation length that is smaller than the size of the system but still grater than $l$ when $\gamma = 0.0$, indicating the onset of non-trivial rearrangements. Here, non-trivial rearrangements refer to those that result from an interplay between the field and droplet-droplet interactions. Such dynamics can be characterized by fluctuations $\chi$ in the order parameter:

\begin{equation}
    \chi = \langle z^2 \rangle - \langle z \rangle^2
\end{equation}

\noindent
where we see a peak at intermediate values of $\gamma$ indicating substantial fluctuations of the individual degrees of freedom from the average behavior of the system. These fluctuations arise from phase differences between neighboring droplets that will eventually lead to doublets and in some cases triplets. Through the additional energy injection of driving, we can see that there is an enhancement of ~ 70 \% in lower energy triplet structures compared with the rapid quench (\textbf{Figure 4d}). This increase in triplets appears to come from the partial conversion of higher energy doublet structures. Comparison with our numerical model ($\xi = 4$) gives good agreement in terms of the number of triplets observed. Interestingly, the number of doublets observed in the numerical annealing experiment is nearly zero indicating a complete conversion of doublets in comparison with the numerical rapid quench. This means that, in simulation, all doublets lie along a reaction coordinate that is both energetically and kinetically accessible due to the annealing field. This is not the case in experiment. It is possible that a slower annealing period may have led to a higher rate of conversion, since this would be closer to an adiabatic limit where $Pe \sim 0$.

\subsection*{Discussion}

We have presented a method to visualize complex annealing dynamics in a many-body system, using motile, binary droplets on a patterned surface. This represents the first numerical and experimental investigation concerning the behavior of ensembles of evaporating motile droplets with a capacity for self-assembly. While the mobility of individual droplets has been extensively studied from both fundamental and applied perspectives, their ability to self-assemble had previously not been realized, providing a new connection between droplet science and  self-assembly \cite{grzybowski2004, volpe2020}. We show that much of the underlying complexity associated with vapor-mediated interactions and Marangoni contraction can be captured by a simple force-based mathematical description, many aspects remain to be more fully explored. For example, the evolution of the vapor field and the time scale associated with signal propagation are clearly important for understanding many-body dynamics of the system but were only treated implicitly. Furthermore, recent experimental work on Marangoni contracted droplets points the way towards a more diverse set of interactions to be explored. For example, gravitational effects on Marangoni flows could lead to repulsive interactions \cite{lohse2019}. Alternatively, expanding the set of evaporating species could lead to multiple, chemically distinct vapor gradients \cite{cira2015}. The system presented here allows for the physical and chemical diversity possible in motile droplets to be studied in a context of condensed matter physics.

While the physical nature of the system limits direct comparisons with electronic or gravitational systems, identifying universal features associated with long-ranged interacting many-body systems ($\sim 1/r$) is nonetheless of fundamental interest. The flexibility of the platform enables us to explore many-body interactions on not just periodically repeating lattices but arbitrary graphs. This flexibility combined with the ability to characterize the system in terms of Kuramoto coherence invites table top experimental investigations into synchronization phenomena where frustration and spatial arrangement are important, for example in neural circuits and electrical grids \cite{munoz2014, pedersen2008}. Meanwhile, the ability to directly associate a non-equilibrium driving force with the resulting micro trajectory provides new opportunities to develop control strategies for systems far from equilibrium in an experimental context. While some control strategies have been proposed for coupled oscillators described by Kuramoto dynamics \cite{kori2007}, recent progress in the use of artificial intelligence to gain insight into complex systems like glasses and proteins extends the limits of what is possible \cite{alphaFold2020, bapst2020, sood2020}. We expect that the slow timescales and ease of observation makes this system an ideal playground to develop inference strategies for the robotic control of complex systems enabling novel manufacturing methods \cite{mnih2015, abbeel2018}. An open challenge remains to test the limits of state control in various geometries subject to arbitrary time-dependent global fields. Furthermore, the discrete nature of this system could prove a powerful tool for inspiring and validating theory in non-equilibrium statistical mechanics, particularly understanding small and disordered systems\cite{england2015}. With a growing capability to both observe and manipulate self-assembly systems at the nanoscale there will be an increasing need for experimentally verified theoretical frameworks and practical control strategies\cite{sood2020, tang2016, aliprandi2015, fukui2016}. 

In summary, we have demonstrated that binary droplets can be combined with hydrophobic patterning to produce a rich playground where template-based assembly can be controlled using time-dependent fields. This system gives experimental access to uncharted building blocks of complex systems using only a small number of commonly available materials. The simplicity of this system makes it accessible outside the laboratory, empowering a far broader group of people to explore the complexity that lies at the intersection of geometry and non-equilibrium phenomena.

\section*{Acknowledgments}
We acknowledge all members of the Prakash Lab for useful and exciting discussions, in particular M. S. Bull. A. R. M. is supported by the National Science Foundation Graduate Research Fellowship Program. M.P is supported by NSF Career Award, Keck Foundation award, HHMI-Gates Faculty Scholar Award, CZI Biohub Investigator Award and NSF grant DBI-1548297. We acknowledge A. W. Lei for helping during initial set of experiments. We acknowledge H. Li for help with imaging. We further acknowledge T. Pollina for thoughtful discussions concerning the construction of the rotary actuated Stewart Platform. 

\bibliographystyle{naturemag}
\bibliography{references}

\newpage

\begin{figure*}
\includegraphics[width=0.5\textwidth]{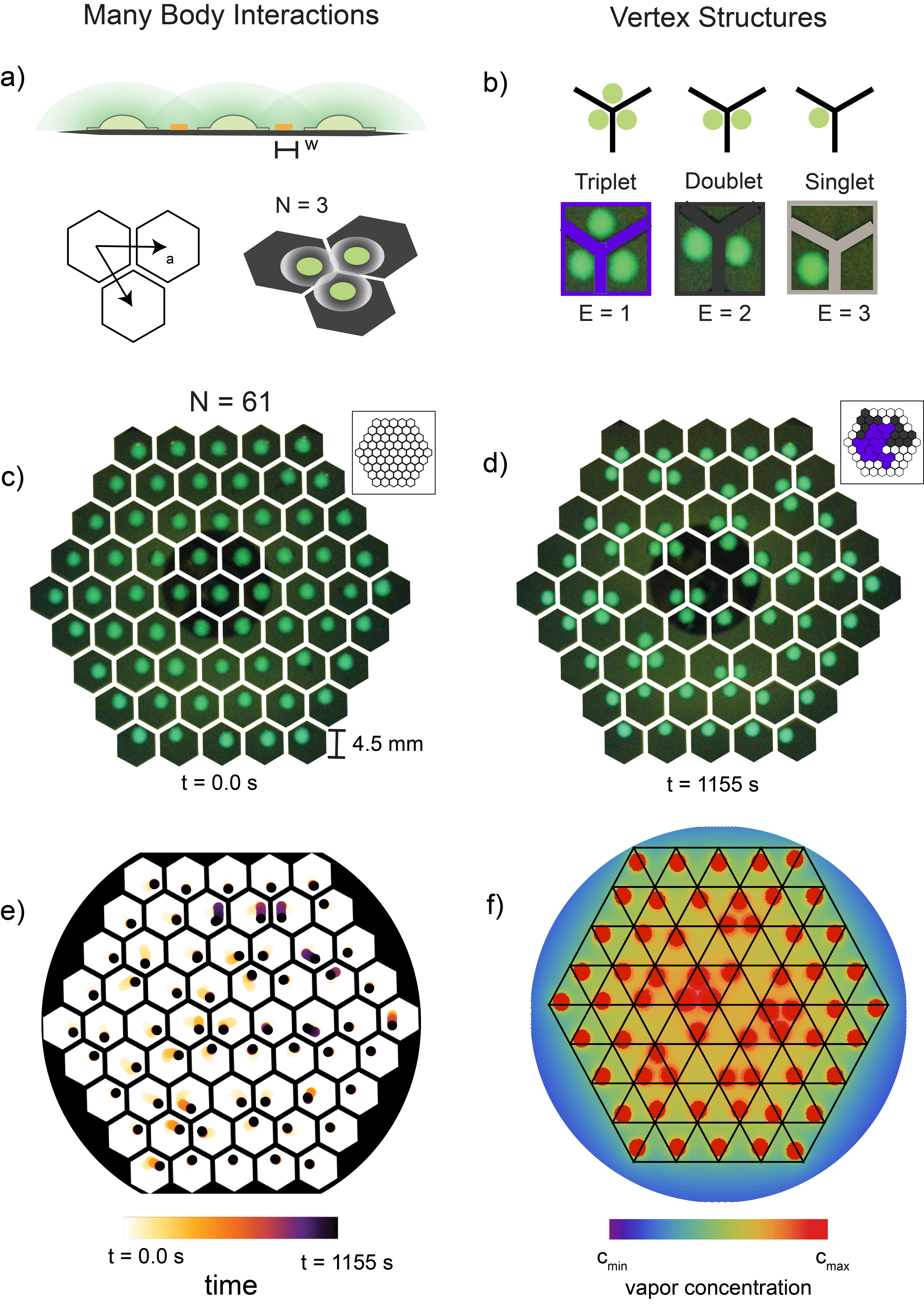}
\caption{\label{Fig1} 
\textbf{Long-ranged many-body interactions on a lattice}. 
\textbf{a-top}, A hydrophobic barrier of dimension $ w $ can be used to confine binary marangoni-contracted droplets into lattice sites. Vapor gradients are unaffected by these obstacles allowing, droplets to interact with their neighbors over long distances. 
\textbf{a-bottom}, Here, we study droplets interacting on a hexagonal honeycomb lattice characterized by lattice vector $ a $. In a $ N = 3 $ lattice, we expect the droplets to deterministically organize at a vertex.
\textbf{b}, However, Extending the size of the lattice to higher $N$ results in a frustrated system where there are multiple types of vertex structures that can form characterized by droplet occupany ranging from 0 to 3.
\textbf{c-d}, Representative time series of an experimental realization of the system with $N=61$ lattice sites, insets show vertex structures at initial \textbf{c} and final \textbf{d} states color coded according to the scheme in \textbf{b}.
\textbf{e}, We can observe the microscopic dynamics associated with each lattice site in response to a complex vapor field.
\textbf{f}, Numerical model of complex vapor field with dual lattice overlaid, showing that vertices in dual-space correspond to potential sites of vapor enrichment.
}
\end{figure*}

\begin{figure*}
\includegraphics[width=0.95\textwidth]{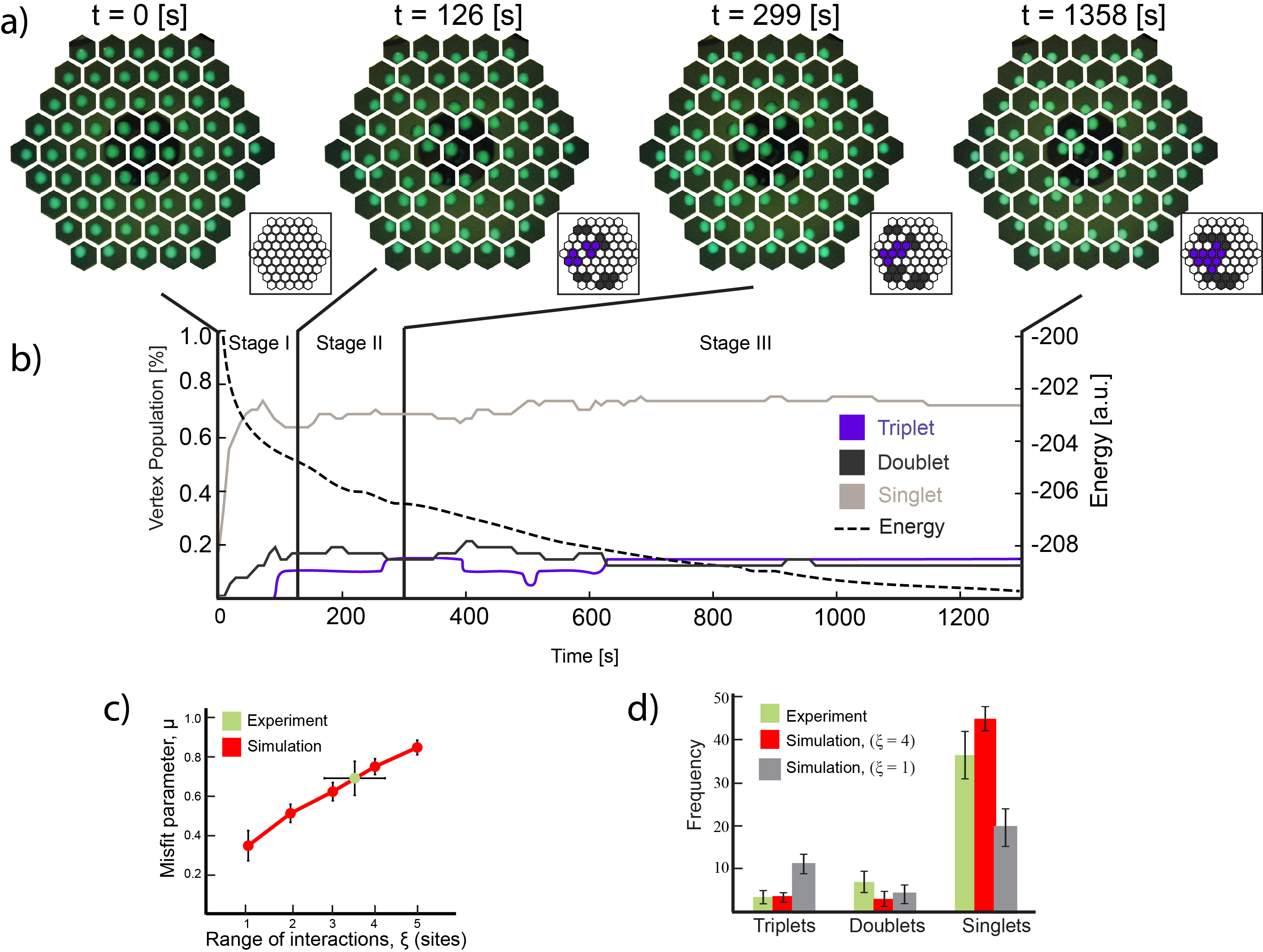}
\caption{\label{Fig2} 
\textbf{Rapid quench from a high energy state}. 
\textbf{a}, Representative time series showing that the droplet relaxation process is characterized by three stages. First, a high energy state is initiated by placing each droplet in the center of its unit cell. Stage I is characterized by a rapid, radially-inward collapse until droplets encounter a hydrophobic boundary. Stage II describes droplets moving towards well-defined vertex states from their first point of contact with the hydrophobic boundary. Finally, stage III describes rearrangements over longer timescales.
\textbf{b}, Monitoring the population of vertex sites as a function of time shows that doublets and singlets form first and are the dominant structure while triplets form later and often from the conversion of doublets. The energy during this process remains a monotonically decreasing function of time consistent with the analogy that this process is a rapid quench. 
\textbf{c}, Misfit parameter plotted as a function of number of interacting neighbors $\xi$ comparing both simulation and experiment, revealing that experiment agrees with a discrete value in the range $\xi = 3-4$.
\textbf{d}, Comparison of ($N=33$) experiments with numerical simulations shows that $\xi = 4$ neighbors provides a better description of the coarse-grained properties of the system compared with simulations considering only nearest-neighbor ($\xi = 1)$ interactions.
}
\end{figure*}

\begin{figure*}
\includegraphics[width=\textwidth]{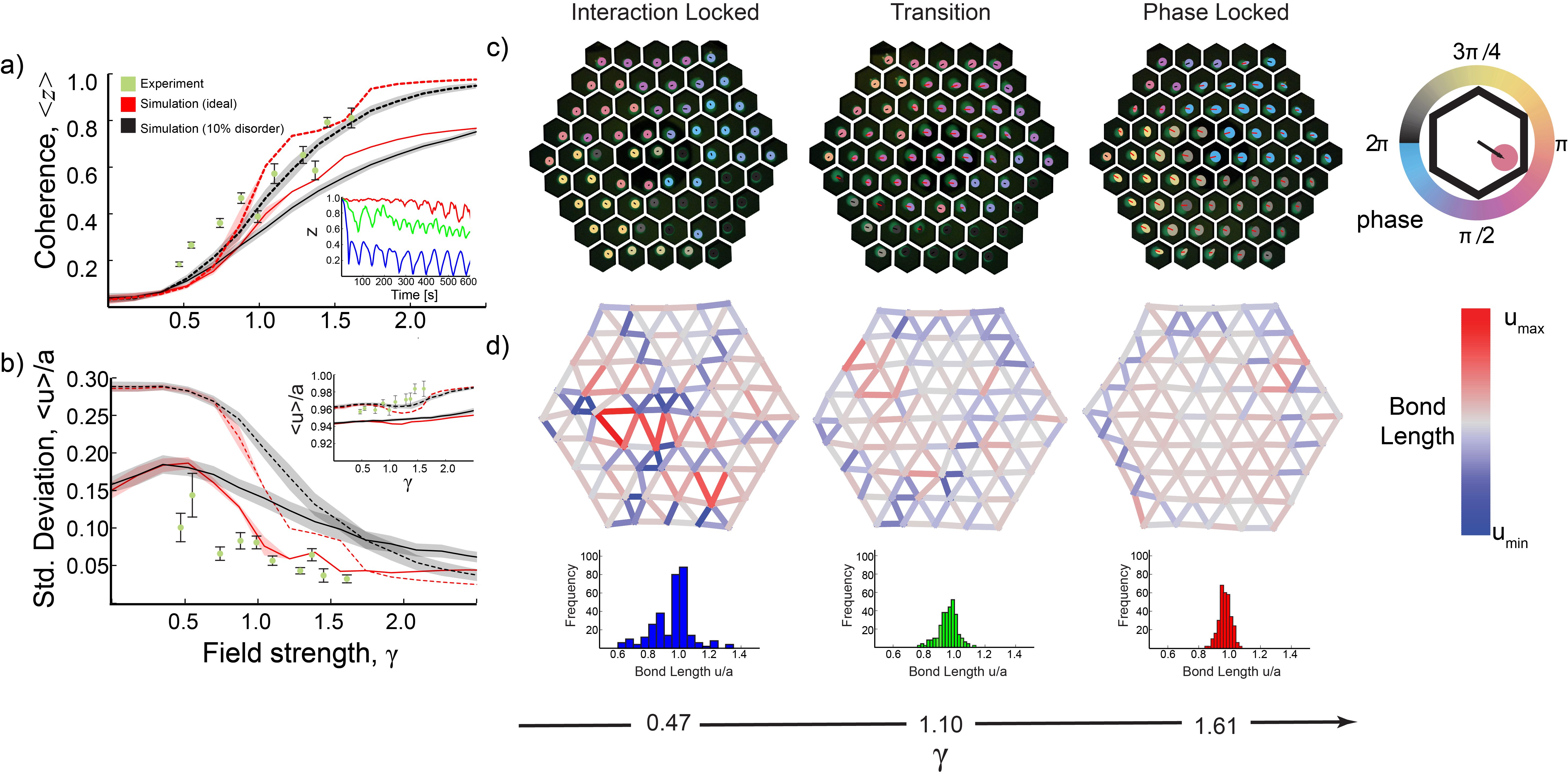}
\caption{\label{Fig3} 
\textbf{Field-driven symmetry breaking}. 
\textbf{a}, A second-order like phase transition is observed in the Kuramoto order parameter $\langle z \rangle$ as a function of field strength $\gamma$. Experimental data is compared with numerical simulation for $\xi = 1$ (dashed line) and $4$ (solid line) neighbors with 0\% and 10\% disorder in droplet size. Inset shows time dependence of $z$ for $\gamma = 0.47, 1.10$, and $1.61$ depicted in blue, green, and red, respectively.
\textbf{b}, Visualization of the microscopic details of the system highlights the spatial dependence of droplet trajectories and phase behavior throughout the symmetry breaking process. Mobile and stationary droplets are distinguished by red and black lines, respectively. 
\textbf{c}, This symmetry breaking can also be characterized in terms of the standard deviation in period-averaged bond lengths $\langle u\rangle$ (inset).
\textbf{d}, Spatial dependence of $\langle u \rangle$ for the data shown in \textbf{b} with corresponding histogram for individual $u$.
All error bars represent one standard deviation of period-to-period variation.
}
\end{figure*}

\begin{figure*}
\includegraphics[width=\textwidth]{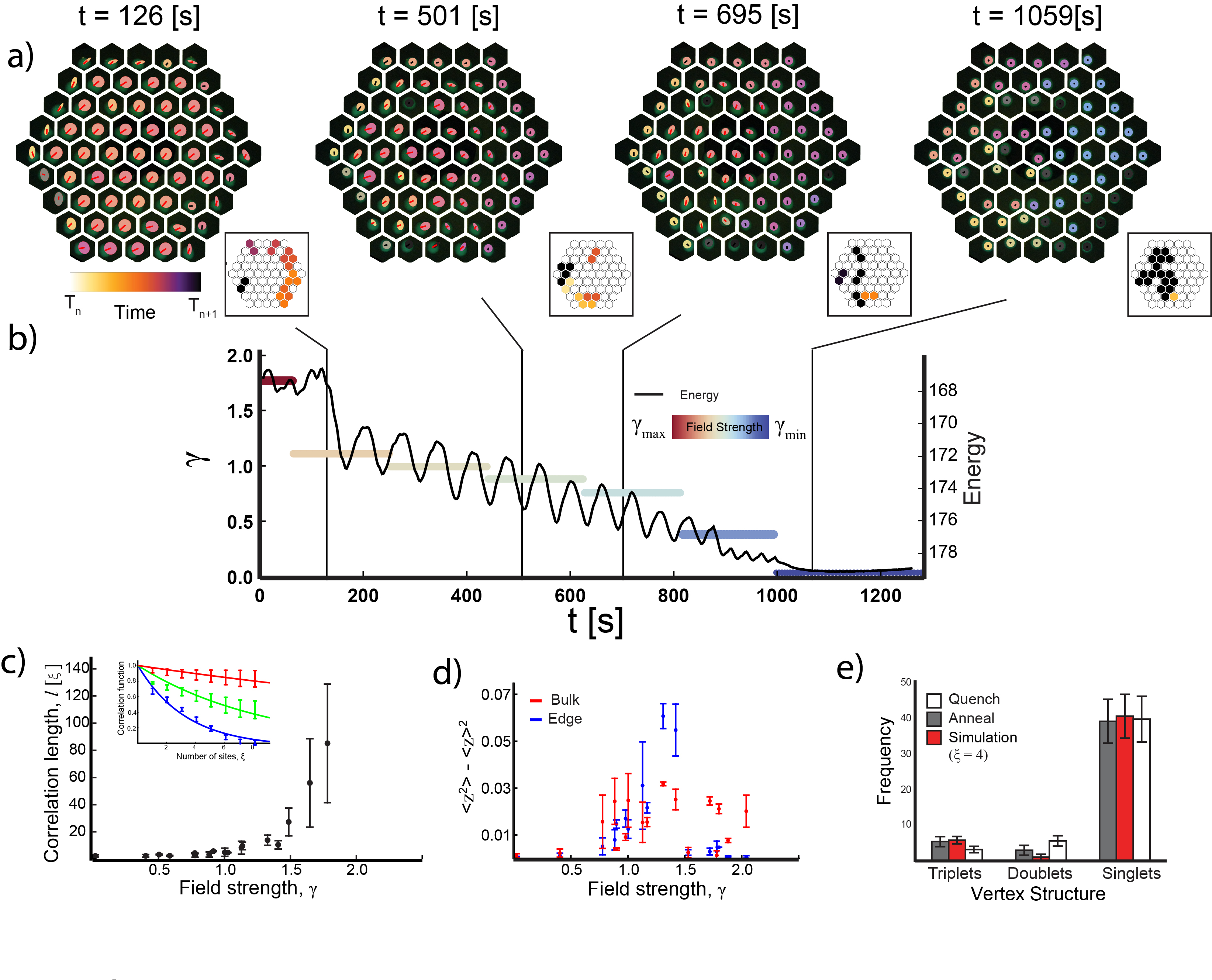}
\caption{\label{Fig4} 
\textbf{Field-induced annealing}
\textbf{a}, Representative time series showing the droplets subject to an annealing protocol, driving the system from a high to low energy state with the phase and trajectory of individual droplets indicated by a colored ellipse. Color code is the same as in \textbf{Figure 3}. Insets show lattice sites that form transient vertex structures with the color of the icon corresponding to the time at which the structure was broken into singlets. Black hexagons correspond to structures that persisted into the next period or until the end of the experiment if the drive was sufficiently low ($t = 1059$ $s$).
\textbf{b}, Anneal schedule consists of a sequence of values for the field control parameter $\gamma$ (strength indicated by color) as a function of time. The schedule shown here begins in a phase locked state and then proceeds to slowly reduce $\gamma$ around the transition region. The energy of a typical annealing experiment is overlayed (black line) showing that - unlike the rapid quench - the system makes periodic excursions to higher energy states consistent with the analogy to annealing. 
\textbf{c}, As the influence of the field on the system is increased, diverging correlation lengths are observed.
Inset shows correlation function as a function of distance in terms of lattice sites where $\gamma = 0.47, 1.10$, and $1.61$ are depicted in blue, green, and red, respectively.
Intermediate field strengths (green) result in correlation lengths that are on the order of system size.
\textbf{d}, At these intermediate field strengths, fluctuations in the order parameter are high with edge droplets experiencing higher fluctuations than those in the bulk.
\textbf{e}, Comparison of annealed experiments ($n=12$) with  simulations implementing the numerical equivalent annealing schedule with $\xi=4$ interactions showing good agreement in terms of increased number of low energy triplet states observed compared with rapid quench experiments ($n=33$).
}
\end{figure*}

\end{document}

\title{Supplementary Information - Evaporative self-assembly of motile droplets}

\author{Anton Molina}
\affiliation{Department of Materials Science and Engineering, Stanford University, 496 Lomita Mall, Stanford, CA 94305, USA}
\affiliation{Department of Bioengineering, Stanford University, 443 Via Ortega, Stanford, CA 94305, USA}

\author{Shailabh Kumar}
\affiliation{Department of Bioengineering, Stanford University, 443 Via Ortega, Stanford, CA 94305, USA}

\author{Stefan Karpitschka}
\altaffiliation[Current Address: ]{Max Planck Institute for Dynamics and Self-Organization (MPIDS),Am Fa{\ss}berg 17, 3077 G{\"o}ttingen, Germany}
\affiliation{Department of Bioengineering, Stanford University, 443 Via Ortega, Stanford, CA 94305, USA}

\author{Manu Prakash}
\email[Correspondence: ]{manup@stanford.edu}
\affiliation{Department of Bioengineering, Stanford University, 443 Via Ortega, Stanford, CA 94305, USA}

\date{\today}
\maketitle

\makeatother
\setcounter{figure}{0}
\setcounter{section}{0}
\setcounter{table}{0}
\setcounter{page}{1}

\tableofcontents

\section{Supplementary Methods}
%\label{sec:materials_methods}

%\subsection{Materials}
%Water, Propylene Glycol, Opticz (Direct Glow, USA)

\subsection{Fabrication of lattices}

\subsubsection{Photolithography}

Single-side polished, silicon (Si) wafers (diameter: 4 inch, thickness: 550 $\mu$m) were obtained from University Wafers (USA). Microposit S1813 G2 positive photoresist (DOW Inc., USA) was spuncoat onto the wafers with target photoresist thickness 1.5 $\mu$m. The wafers were softbaked at 60 $^{\circ}$C for 1 minute. These wafers were then exposed on a Quintel mask aligner (Neutronix, USA) using photomasks (CAD/Art Services Inc., USA) with desired lattice features. Photoexposed Si wafers were developed for 30 seconds using Microposit MF 319 developer (Shipley company, USA), washed with water and dried. Electron-beam deposition was used to coat the wafers with a thin layer of chromium (thickness: 5 nm), which serves as an adhesion layer and gold (thickness: 100 nm) as the top layer. The wafers were then placed in acetone for metal lift-off: photoresist still remaining on the wafers was dissolved in acetone, along with the removal of overlying metal layers. Wafers were then cleaned with deionized water and dried. Si wafers with thin-metal lattice features were obtained at the end of the process. 

Before an experiment, lithographically patterned wafers are treated with air plasma for 30 s and then immediately immersed in a 5 mM dodecanethiol in 70 \% ethanol.  The thiolated surfaces are rinsed with ethanol and briefly sonicated (approx. 10 s) to remove any physisorbed molecules. Experiments are initiated 30 minutes after activation to avoid any effects associated with contact angle recovery\cite{Yamamoto2019}. Wafers can be reused multiple times with appropriate cleaning. Wafers are cleaned by exposing them to air plasma for 20 min, sonicating for 10 min in acetone, rinsing with $70\%$ ethanol, and finally drying in a nitrogen stream. 

\subsubsection{Ink-based fabrication}
Large-scale, high-quality hydrophobic lattice patterns can be obtained using multi-surface permanent markers (Sharpie\textsuperscript{TM}) by programming an automated plotter/cutter (Roland Camm1-Servo) with CAD tools. We also developed a frugal method for developing lattice patterns that requires the use of only a rubber stamp and a multi-surface ink pad (StazeOn). Custom rubber stamps can be designed using commonly available software and obtained from online retailers at a low-cost. We note that rubber stamps tend to produce noisier patterns, characterized by rough edges.

\subsection{Experimental design}

Binary droplets consisting of water/propylene glycol (PG) mixtures were used with a composition indicated by the $\%$ PG. Unless, otherwise indicated experiments were carried out using $30\%$ PG. This value was chosen as a balance between droplet lifetime which is reduced with increased $\%$ PG and droplet mobility which increases with reduced $\%$ PG. The droplets are visualized by adding a small amount (0.08 $\%$) $[V/V]$ of commerically available UV active dye (Opticz, DirectGlow.com) that comes dissolved in isopropanol. For experiments with a small number of droplets ($N < 7$), a micropippete was used to deposit the droplets. For experiments with a larger number of droplets, the droplet solution is dispensed into a polystyrene petri dish and a custom built tool is used to deposit the droplets in well-defined configurations in parallel. 

Experiments were carried out in an enclosed chamber that minimized any disturbances to the droplets from ambient air currents. The droplet array is illuminated at a shallow angle with UV light ($\lambda = 400$ nm) and imaged at 1 FPS from a USB camera (See3CAM 130,eCon Systems) with a long pass filter cube ($\lambda = 495$ nm) placed above the lattice. A DHT-22 sensor (Adafruit Industries, USA) is placed in the chamber to measure the humidity and temperature at the boundary of the system. We note that the illumination did not lead to any significant increase in temperature during experiment (SI Figure 1a). This chamber sits atop a custom-built, rotary-actuated Stewart platform with 6 degrees of freedom \cite{stewart1965, patel2017, szufnarowski2013}. This design was chosen for its ability to support a large chamber with a high level of stability. Briefly, the set up consists of 6 servo motors (Dynamixel AX-12A, Robotis) controlled by an Arduino compatibile microcontroller (Arbotix-M Robocontroller, Trossen Robotics). The servos labelled $i=1-6$ and are arranged in 3 pairs separated by $120 \degree$.

To coordinate the motion of the 6 servo motors so as to produce a precessing gravity vector, each of the $n$ pairs is given a phase offset of $2n\pi/3$.  The motion of the platform can then be controlled with a specified amplitude $A$ and period $T$ to give an angle update $\beta$

\begin{equation}
    \beta =A \cos{(\frac{2 \pi t}{T} + \frac{2n\pi}{3})}
\end{equation}

\noindent
This update is converted into microseconds and sent to the $i^{th}$ motor which is incremented depending on its location:

\begin{equation}
    \beta^{(i)}_{t+1} =
    \begin{cases} 
      \beta^{i}_{t} + \beta  & i, even \\
      \beta^{i}_{t} - \beta  & i, odd
   \end{cases}
\end{equation}

\noindent
The orientation of the platform is measured in two ways. First a bulls-eye spirit level is used as a quick check to ensure that the platform is level for static experiments. Second, a BNO-055 digital accelerometer (Bosch Sensortec GmbH, Germany) is used to measure the time evolution of the field during experiments. The sensor is configured to report quaternion values $(w, x, y, z)$ which can be used to construct a three dimension rotation matrix for every time step: 

\begin{equation}
    R = 
    \begin{pmatrix}
    \begin{smallmatrix}
    w^2+x^2+y^2+z^2 & 2(xy-wz) & 2(wy+xz)\\
    2(xy+wz) & w^2-x^2+y^2-z^2 & 2(-wx+yz)\\
    2(-wy+xz) & 2(wx+yz) & w^2-x^2-y^2+z^2
    \end{smallmatrix}
    \end{pmatrix}
\end{equation}

\noindent
The matrix $R$ can be used to obtain an orientation vector $\textbf{v}$ by through multiplication of the unit vector normal $\textbf{n} =  [0, 0, 1]^T$ to the platform surface in a flat configuration. The tilt angle $\alpha$ between this normal vector and the orientation vector is calculated according to:

\begin{equation}
    \alpha = \arccos{
    \frac
    {
    \textbf{v}\cdot\textbf{n}
    }
    {
    \norm{\textbf{v}}
    \norm{\textbf{n}}
    }
    }
\end{equation}

\noindent
The angle used in calculating the non-dimensional control parameter $\gamma$ (described below) is obtained by taking the average of $\alpha$ over the time series. Two dimensional projections of the field are obtained by taking the x and y components of $\textbf{v}$. Examination of these projections for increasing amplitudes shows the emergence of an asymmetry in the gravity field reflecting the underlying geometry of the platform. However, at these amplitude, the gravity vector is sufficiently large that it is not reflected in the droplet trajectories.

\subsubsection{Rapid quench experiments}
Relaxation experiments are initialized with all droplets approximately at the center of their unit cell. The wafer is then placed into the reaction chamber observed for a maximum of 21 minutes, corresponding to a $5 \%$ change in diameter and a narrowing of their distribution (\textbf{SI Figure 2}). The experiments analyzed in this study were carried out at relative humidity of $65 \pm 2.3 \%$ and a temperature of $23 \pm 0.4 \degree C$. The time evolution of humidity and temperature throughout the experiment are shown in \textbf{SI Figure 1}.

\subsubsection{Triplet force balance experiments}
The force experienced by a droplet held in a triplet configuration was measured by seeding three droplets with volume $V = 2 \mu L$ into a triplet configuration like that depicted in \textbf{SI Figure 3a} separated by a hydrophobic boundary of width $w = 1$ $mm$. The droplets were then raised to an angle $\alpha$ relative to the horizontal and observed for up to $100$ $s$. The orientation of the gravity vector was directed along the "weak" axis as shown in \textbf{SI Figure 3b}. If the triplet cluster was broken within 20 seconds, the associated angle was categorized as sufficient to break the structure , allowing for comparison with numerical simulations for an equivalent droplet configuration (\textbf{SI Figure 3c-d}). This comparison is basis for computing the non-dimension field parameter $\gamma$ defined as $\gamma = \frac{sin(\alpha)}{sin{(\alpha_{c})}}$ Here, we experimentally determined $\alpha_{c} = 3.5 \degree$ which is in good agreement with theory (see below)

\subsubsection{Field-driven experiments}
Constant field-driven experiments are initialized with all droplets located at the vertex corresponding to the orientation of the gravity vector at $t=0$. Experiments are performed over a range of angles $\alpha=0-7.9 \degree$ for a fixed period $T=60$ s. Each experiment is observed for 10 periods. The 2D projection of the precessing gravity vector for the different field strengths used in this experiment are shown in \textbf{SI Figure 5}. The average temperature and humidity for these experiments are $22 \pm 0.3 \degree C$ and $60 \pm 3.3 \%$, respectively.

\subsubsection{Field-induced annealing experiments}
Annealing experiments are initialized with droplets in the same configuration used in the driven experiments. Experiments begin with $\alpha = 8.3 \degree$ and reduced to zero following the annealing schedule shown in \textbf{Figure 4}, comprising 15 periods at $T = 60$ s followed by 5 minutes at $\alpha = 0 \degree$. Thus the experiment has a total duration of 20 minutes, providing comparison with rapid quench experiments. The experiment was repeated 12 times with temperature and humidity at $24 \pm 0.3 \degree C$ and $54 \pm 5.3 \%$, respectively.

\subsection{Image processing}
Images collected from experiment were cropped using the ImageJ software package to include only the part of the frame containing the lattice. Subsequent processing was done using Mathematica. First, the droplets were segmented by binarizing the green color channel with a threshold value of 0.55-0.60. Next, a hexagonal mask is made to match the lattice shape and is subtracted from the binarized image. This has the advantage of removing any noise in segmentation from the lattice and to unambiguously associate each droplet with the correct unit cell. Finally, Mathematica's built-in SelectComponents function isolated each droplet and the built-in ComponentMeasurements function calculates the location of every droplet centroid and diameter. An empirical conversion factor of 10.6 pixel per mm was calculated allowing droplet size to be measured in physical units.

\section*{Supplementary Discussion}

\subsection{Evaluation of interaction force}
\label{sec:droplet_attraction}

Here we follow the analysis given by Cira \textit{et al}. When placed on a high energy surface, two phase marangoni contracted droplets have an apparent contact angle $\theta_{app}$ that is maintained by an evaporation induced surface tension gradient. If there is a nonuniform vapor field, the droplet will move in the direction of higher vapor concentrations. The force a single droplet feels in a non-uniform vapor field can be calculated by integrating the surface tension along the perimeter of a droplet:

\begin{equation} \label{eq:1}
    F_{a} = \int_{0}^{2 \pi} \gamma_{film} R d \theta
\end{equation}

\noindent
Assuming a sudden difference in surface tension between the bulk of the droplet $\gamma_{bulk}$ and the thin film allows $\gamma_{thin}$ to be expressed in terms of measureable quantities $\theta_{app}$ and  $\gamma_{bulk}$ through  a quasi-static horizontal force balance $\gamma_{film}=\gamma_{bulk} \cos{\theta_{app}}$:

\begin{equation}
    F_{a} = \int_{0}^{2 \pi} \gamma_{bulk} \cos{\theta_{app}} R d \theta
\end{equation}

\noindent
Measurement of $\theta_{app}$ of isolated droplets at different values of relative humidity $\phi$ led to an empirical relationship between $\theta_{app}$ and $\phi$:

\begin{equation}
    \cos{\theta_{app}} = m \phi + b 
\end{equation}

\noindent
If we assume a spherical source of humidity at an angle $\theta$ at $r > R$ with radius $R_{s}$ at steady state, we can use the heat equation $D \nabla^2 \phi(r) = 0$ with boundary conditions $\phi(R)=1$ and $\phi(\infty)=\phi_{\infty}$ to obtain:

\begin{equation}
    \phi(r) = \frac
    {(1-\phi_{\infty})R_{s}}
    {r(\theta)}+\phi_{\infty}
\end{equation}

\noindent
It is important to recognize that interaction length is a function of $\theta$ and can be obtained by considering the geometry of the problem:

\begin{equation}
    r(\theta) = \sqrt{(d + R_{s} \cos{\theta})^2 + (R_{s} \sin{\theta})^2}
\end{equation}

\noindent
Making these substitutions and recognizing from symmetry that only the x component contributed to motion allows us to rewrite the attractive force as:

\begin{equation}
    F_{a} = 2 \gamma_{bulk} m R \int_{0}^{\pi}\frac
    {(1-\phi_{\infty}R_{s}) \cos{\theta}}
    {\sqrt{d^2 + R_{s}^2 + 2dR_{s}\cos{\theta}}}
    d\theta
\end{equation}

\noindent
This is a transcendental equation involving elliptical integrals of the first and second kind. Numerical evaluation shows that the dependence of force on distance to a vapor source goes as ${1}/{r^2}$. Therefore we conclude that droplets response to gradients in the vapor field. Evaluating this expression for droplets interacting across a 1 mm barrier gives approximately $2.9 \mu N$.

\subsection{Evaluation of drive force}

Compared with a typical sessile droplet, binary droplets experience very little surface pinning. A sessile droplet of viscosity $\eta$, surface tension $\gamma$, volume $V$, density $\rho$, and contact angle $\theta$ subject to a gravitational force $F_{g} = \rho V g \sin(\alpha)$ will only move when $F_{g}$ is greater than a threshold pinning force given by $V^{1/3} \gamma \theta \Delta \theta$, where $\alpha$ is the angle of the slope relative to the horizontal and $g$ is the gravitational acceleration. Thus, the capillary number $Ca$ is a linear function of the Bond number $Bo$ times $\sin(\alpha)$ with a non-zero intercept. In contrast, since binary droplets move on a thin film of their own constituents, it has been shown that Bond number is directly proportional to the capillary number: $Bo = Ca \sin(\alpha)$ \cite{benusiglio2018}. This observation allows us to use a simple expression free from pinning thresholds for the droplet response to an applied gravitational force.

Comparison with the magnitude of the attractive force calculated above, allows us to predict the angle $\alpha$ at which the two forces are equivalent. With $\rho = 1000 \frac{kg}{m^3}$ and $V = 2$ $\cdot$ $10^{-9}$ $m^3$ and $F_{a} = 2.9$ $\mu N$ we can calculate:

\begin{equation}
    \alpha_{c} = 
    \arcsin{(
    \frac{F_{a}}{\rho V g}
    )}
\end{equation}

\noindent
We obtain a predicted angle $\alpha_{c} = 6.93 \degree$. However, since the droplets in the "weak" triplet configuration considered here move in tandem \textbf{Figure 3b}, we can treat the droplets as a pair with $V = 4$ $\cdot$ $10^{-9}$ $m^3$ and obtain a predicted angle of $\alpha_{c} = 3.46 \degree$ which is in excellent agreement with our experimental result.

\subsection{Evaluation of Drag Force}
The drag force $F_{drag}$ is proportional to the droplet velocity $U$:

\begin{equation}
    F_{drag} = C_{drag} U.
\end{equation}

\noindent
Previous work by Cira \textit{et al.} has established that the theoretical drag coefficient theory developed for moving three phase contact lines works quite well for the system studied here where the droplet is surrounded by a thin liquid film. The drag coeficient $C_{drag}$ is given by:

\begin{equation}
    C_{drag} = \frac{3 \pi R \eta l_{n}}{\theta}
\end{equation}

\noindent
where $R$ is the droplet radius, $\eta$ is the dynamic viscosity, $\theta$ is the contact angle in radians, and $l_{n}$ is the cutoff constant defined as $l_{n} = ln(\frac{x_{max}}{x_{min}})$ where $x_{max}$ and $_{min}$ are the longest and shortest length scales in the system, respectively \cite{cira2015}. Typically, $x_{max} = R$ while $x_{min}$ is on the order of the molecular size of the liquid, giving a $l_{n}$ values between 10-15. Taking $\eta = 3.43$ $10^{-3}$ $\frac{Pa}{s}$, $\theta = 0.24$, $l_{n} = 11.4$, and typical speeds $U = 1$ $\frac{mm}{s}$, we obtain $F_{drag} = 3$ $N$. This is the same magnitude associated with $F_{a}$, suggesting that dissipative forces are significant. The same conclusion is obtained by considering the Reynolds number ($Re = \frac{\rho R U}{\eta}$). For the same values used above we obtain a value of $Re$ that is of order 1, confirming that the system is in a viscous dissipation regime.

\subsection{Assessment of Cooperative Evaporation Effects}

Following the analysis given by Carrier \textit{et al.}, we can treat a group of droplets as a two component problem with where collection of droplets each with a radius $R$ is regarded as a superdrop with radius $R_{s}$. The first component of the problem involves the super droplet evaporating into the boundary with an effective vapor concentration $\phi_{eff}$ that we assume to be constant. Meanwhile the second component involves the component droplets evaporating into the superdrop each maintaining a vapor concentration $\phi_{s}$. The result of this simple analysis is that compared to $N$ independent droplets, the rate of evaporation is reduced by a factor $f$ given by:

\begin{equation}
    f = \frac{2}{1+J_{0}NR/2R_{s}}
\end{equation}

\noindent
Since the contact angle is low, we approximate $J_0 = 4$. Next, we approximate the radius of the super drop to be 35 mm corresponding to  $R_{s} \sim 18 R$ and take $N=61$. This correspond to a reduction in evaporation rate by a factor of $f \sim 0.25$. We remark that this is a crude approximation and note that this expression underestimates the reduction by a factor of 2 in the original work of Carrier \textit{et al.}. Nonetheless, it does justify longer than expected evaporation rates in our system.

Another relevant consideration for the system studied here is the timescale associated with signal propagation across the lattice. The time $t$ for a molecular of water to diffuse across a distance $x$ is given by Fick's second law:

\begin{equation}
    t = \frac{x^2}{2D}
\end{equation}

\noindent
Taking the system size to be $x \sim 10 cm$ and the mass diffusion constant of water to be $D = 0.242 \frac{cm^2}/{s}$, we determine $t \sim 200 s$.

Previous studies on cooperative evaporation effects have focused on ensembles of sessile droplets \cite{eggers2010, carrier2016, schafle1999, lacasta1998, wray2019, wray2019, khilifi2019}. Coupling motility to droplet evaporation leads to an effective advection component associated with droplet motion. The relative importance of this advective component to pure diffusion can be characterized by the Peclet number $(Pe = \frac{aU}{D})$, where $a$ is the lattice constant, $U$ is droplet velocity. For the speeds observed in the high $\gamma$ regime, $Pe \sim 1$, confirming the importance of advective effects.

\subsection*{Numerical Simulations}

We develop a numerical simulation for the system with droplet radii $R = 2.3 mm$ and lattice constant $a = 7.5 mm$.  Disorder in droplet size was studied by adding a 10\% standard deviation. As discussed above, the system is in a viscous regime, allowing the equations of motion to be represented by a first order system of differential equations \cite{Strogatz}:

\begin{equation}
    \vec{\dot{r}}_{i} = - \zeta \nabla E_{i}
\end{equation}

\noindent
where $E_{i}$ is the energy associated with the $i^{th}$ droplet and $\zeta$ is a damping parameter set to 1. The energy has two terms describing long-range droplet-droplet attractions $E_{a}(\xi)$ and a short range repulsive term due to interactions with the unit cell boundary $E_{cell}$. 

Recalling that the attractive droplet-droplet force $F_{a} \sim 1/r^{2}$, we can identify the associated potential as vapor field established by neighboring droplet $\phi \sim 1/r$. The exact form of $\phi$ is taken to account for the finite size of a droplet with radius R and its saturated vapor field for distances $r < R$ \cite{eggers2010}:

\begin{equation}
    \phi(r_{ij}) = 
    \begin{cases}
    1 & r_{ij} < R \\
    \frac{2}{\pi}\arcsin{(\frac{R}{r_{ij}})} & r_{ij} \geq R
    \end{cases}
\end{equation}

\noindent
This function closely matches the $\sim 1/r$ scaling required to obtain the correct force value. In the context of a lattice the individual $\phi_{i}$ contributions must be summed over, becoming a function of the number of interacting neighbors. Our simulation framework considers energy as a function of the number of neighbors at $\xi$ discrete lattice sites $\mathcal{N(\xi)}$: 

\begin{equation}
    E(\xi) = \sum_{i=1}^{N} \sum_{j = 1}^{\mathcal{N (\xi)}} \phi(r_{ij}) 
\end{equation}

\noindent
The term $E_{cell}$ represents the short ranged repulsive interaction imposed by the hydrophobic material used to fabricate the lattice pattern. Mathematically, this is represented as a logistic function whose parameters are chosen such that the barrier height is approximately an order of magnitude greater than the $E_{a}$ terms:

\begin{equation}
    E_{cell} = \frac{-R}{1+e^{\frac{0.48-R-z(x,y)}{0.023}}}
\end{equation}

\noindent
Here, $z(x,y)$ reflects the unit cells hexagonal symmetry and determines the influence of the boundary given droplets Cartesian coordinates $(x,y)$:

\begin{equation}
    z = \frac{\sqrt{x^2 + y^2}}{c_{0} + \sum^{4}_{n}c_n\cos{(6n \kappa)}}
\end{equation}

\noindent
where $\kappa$ is given by:

\begin{equation}
    \kappa = \arctan{\frac{y}{x}}
\end{equation}

\noindent
the denonminator is a fourier expansion in powers of 6 with coefficients $ c_{n} $ chosen such that derivatives with respect to $ \zeta $ are set equal to zero.

Since the Reynolds number of this system is low $Re \sim 0.1$, we can neglect inertial effects and obtain the dynamics of the system by evaluating:

\begin{equation}
    \vec{\dot{r}}_{i} = - \zeta \nabla E_{i}
\end{equation}

\noindent
Numerical integration was carried out using the NDSolve function in the Mathematica software package (backwards difference method). 

\subsubsection{Static simulations}

Static simulations were carried out to understand the role of number of interacting neighbors $\xi$, varied from 1 to 5. Average values were calculated from $N=1,000$ iterations representing different initial conditions. This value of $N$ was found to be large enough to give reproducible statistics.

\subsubsection{Driven simulations}

Here, we study time varying gravitational fields where the orientation of the field is periodic but the magnitude $A$ is constant over a period. Thus the force can be numerically described by a term: 

\begin{equation}
\label{eq:driving}
    F_{g}(t)
    \begin{cases}
    F_{g}^x(t) = A \sin(\frac{2 \pi t}{T}) \\
    F_{g}^y(t) = A \cos(\frac{2 \pi t}{T})
    \end{cases}
\end{equation}

\noindent
Thus Equation \ref{eq:driving} becomes: 

\begin{equation}
    \vec{\dot{r}}_{i} = - \nabla E_{i} + F_{g}
\end{equation}

\noindent
and can be solved as described above. The simulation had a numerical period of 10 and simulation were run for 10 periods.

\subsubsection{Annealing simulations}

Here, we let the amplitude in Equation \ref{eq:driving} become a function of time. Experimental values for $\alpha$ were used as input to define $A(t)$. Average values were calculated from $N=1,000$ iterations representing different initial conditions. 

\subsection{Supplementary video caption}

\begin{itemize}
    
\item \textbf{Video 1}
Evaporative self-assembly of motile binary droplets with and without external driving.

\item \textbf{Video 2:}
Numerical simulation showing self assembly without external driving with number of interacting neighbors ranging from $\xi = 1-5$.

\item \textbf{Video 3:}
Numerical simulation showing non-equilibrium phase transition characterized by Kuramoto order parameter.

\item \textbf{Video 4:}
Experimental data showing individual droplet phase at three different regimes of the non-equilibrium phase transition.

\item \textbf{Video 5:}
Experimental data showing the nearest neighbor bond-length at three different regimes of the non-equilibrium phase transition.

\item \textbf{Video 6:}
Observation of high energy excursions during an annealing protocol.

\end{itemize}

\bibliographystyle{naturemag}
\bibliography{references}

\newpage

\begin{figure*}
\includegraphics[width=\textwidth]{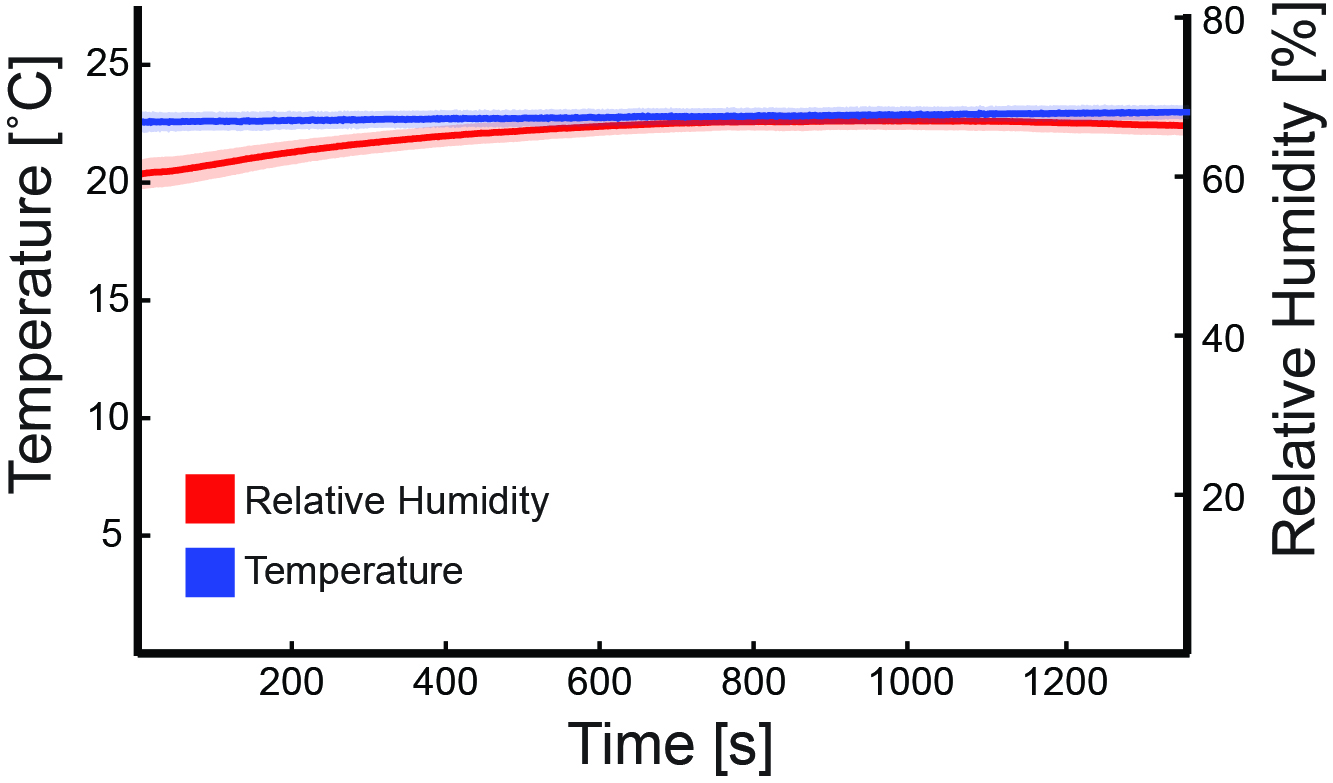}
\caption{\label{SI Fig1} 
Time series data showing evolution of temperature and relative humidity during a typical experiment.
}
\end{figure*}

\begin{figure*}
\includegraphics[width=0.95\textwidth]{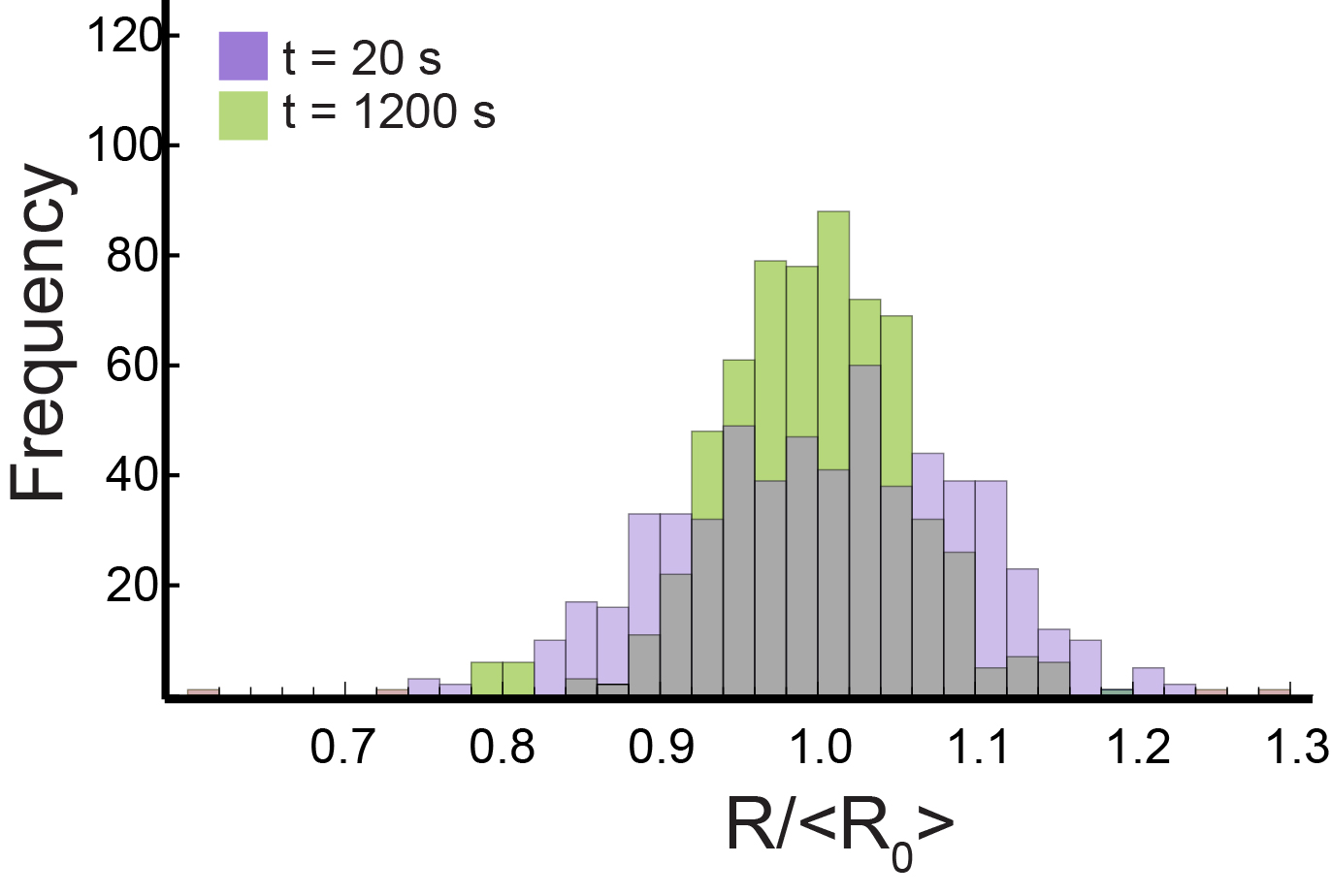}
\caption{\label{SI Fig2} 
Initial and final distribution of droplet radii (N = 610) for experiments with zero driving.
}
\end{figure*}

\begin{figure*}
\includegraphics[width=\textwidth]{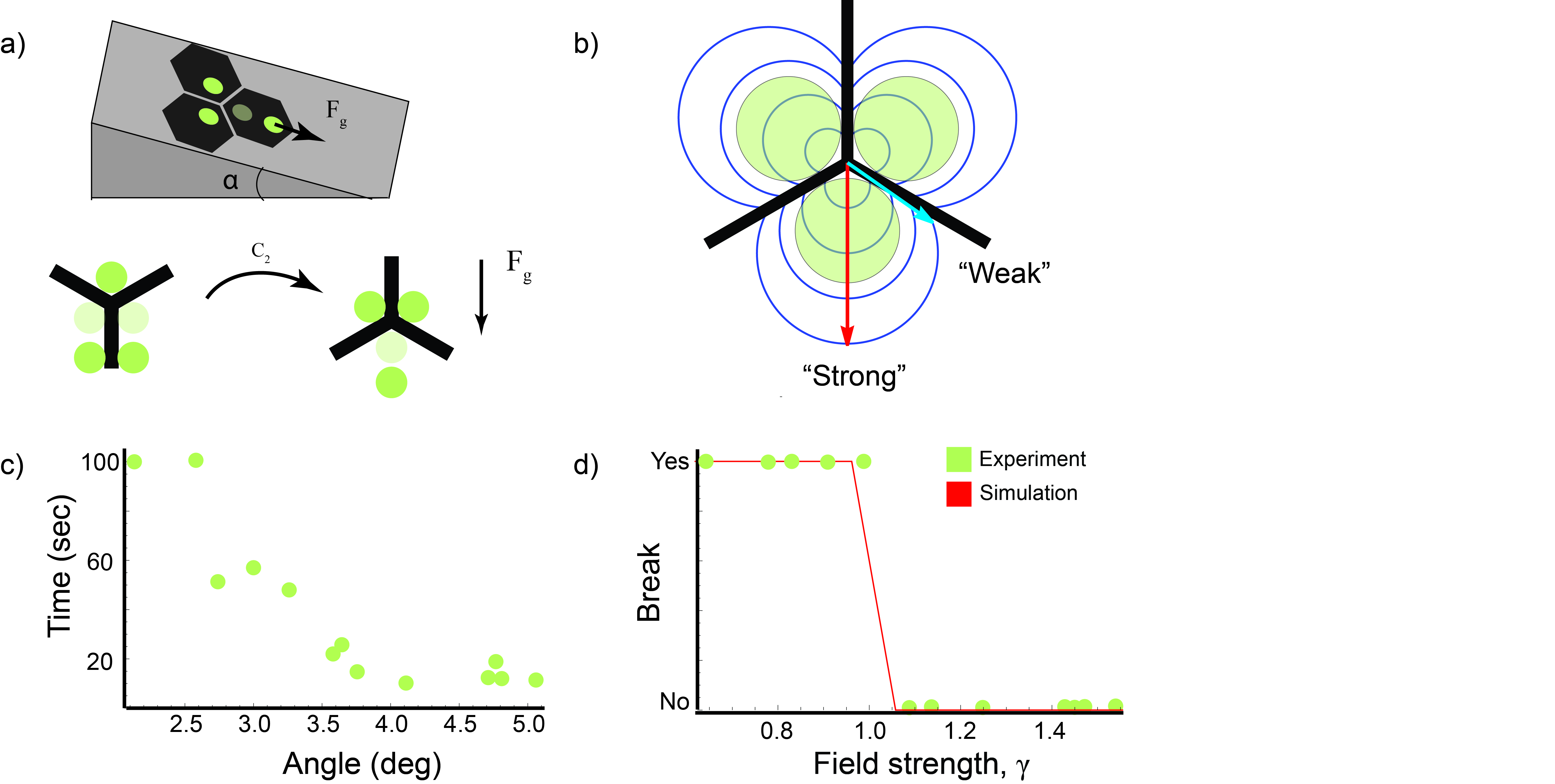}
\caption{\label{SI Fig3} 
\textbf{Force balance measurements for isolated triplets}
\textbf{a}, Cartoon representation of the experiment showing a triplet at two possible orientations $\theta$ defined with respect to the gravity vector and related to each other via a $C_{2}$ symmetry.
\textbf{b}, Contour plot of vapor field overlaid onto a triplet structure showing positions of highest and lowest vapor concentration corresponding to "weak" and "strong" axes where the gravitational force required to break the triplet structure is minimal and maximal, respectively.
\textbf{c}, Experimental measurement of the time required to observe the separation of a triplet into a doublet and singlet as a function of tilt angle.
\textbf{d}, Normalization of data in \textbf{c} using a cutoff criteria of 20 s to define non-dimensional $\gamma$ enabling comparison with equivalent numerical representation.
}
\end{figure*}

\begin{figure*}
\includegraphics[width=\textwidth]{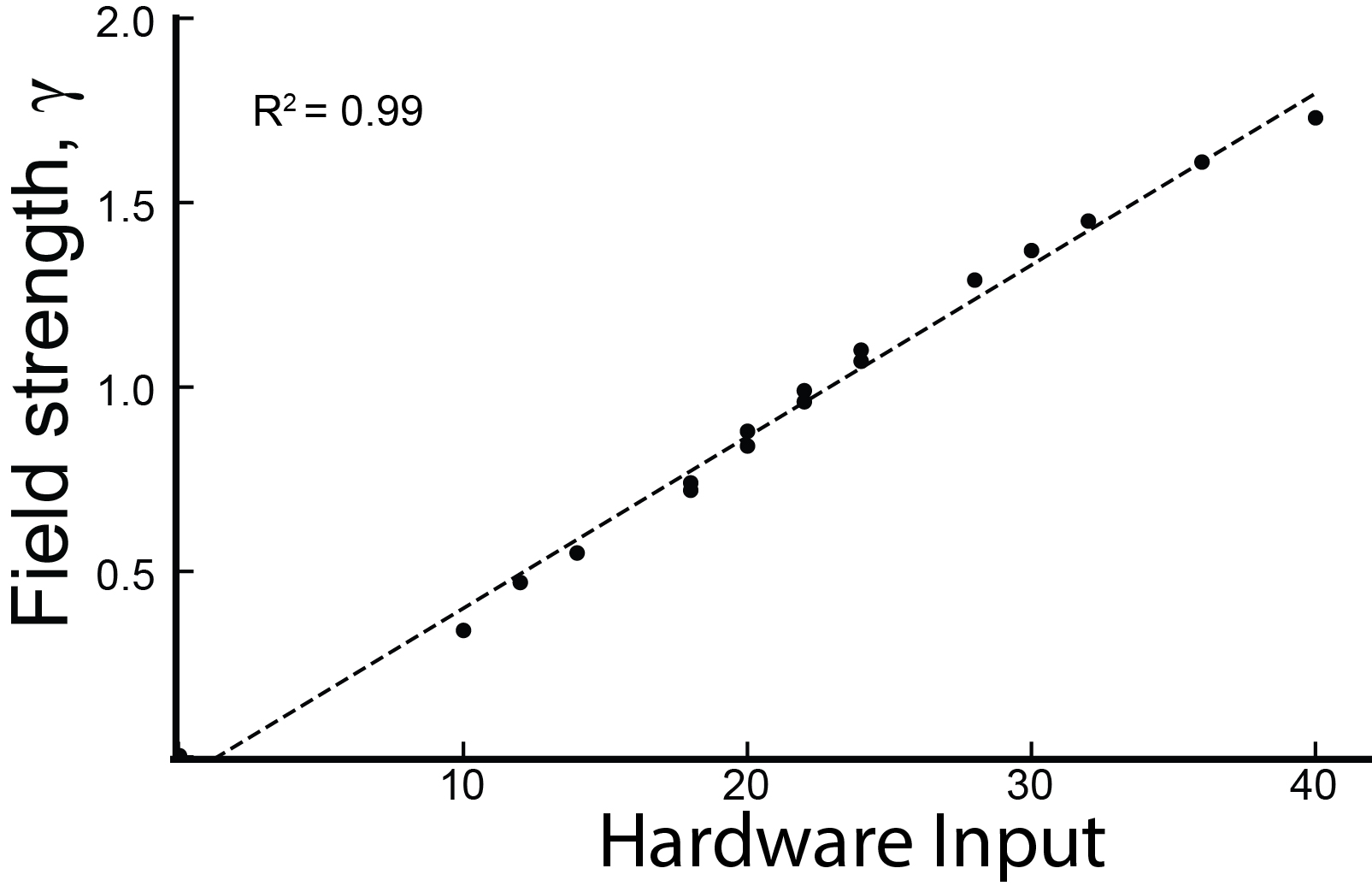}
\caption{\label{SI Fig4} 
Non dimension field parameter $\gamma$ plotted against amplitude used in Arduino microcontroller. Linear fit shows consistency between multiple experiments.
}
\end{figure*}

\newpage

\begin{figure*}
\includegraphics[width=\textwidth]{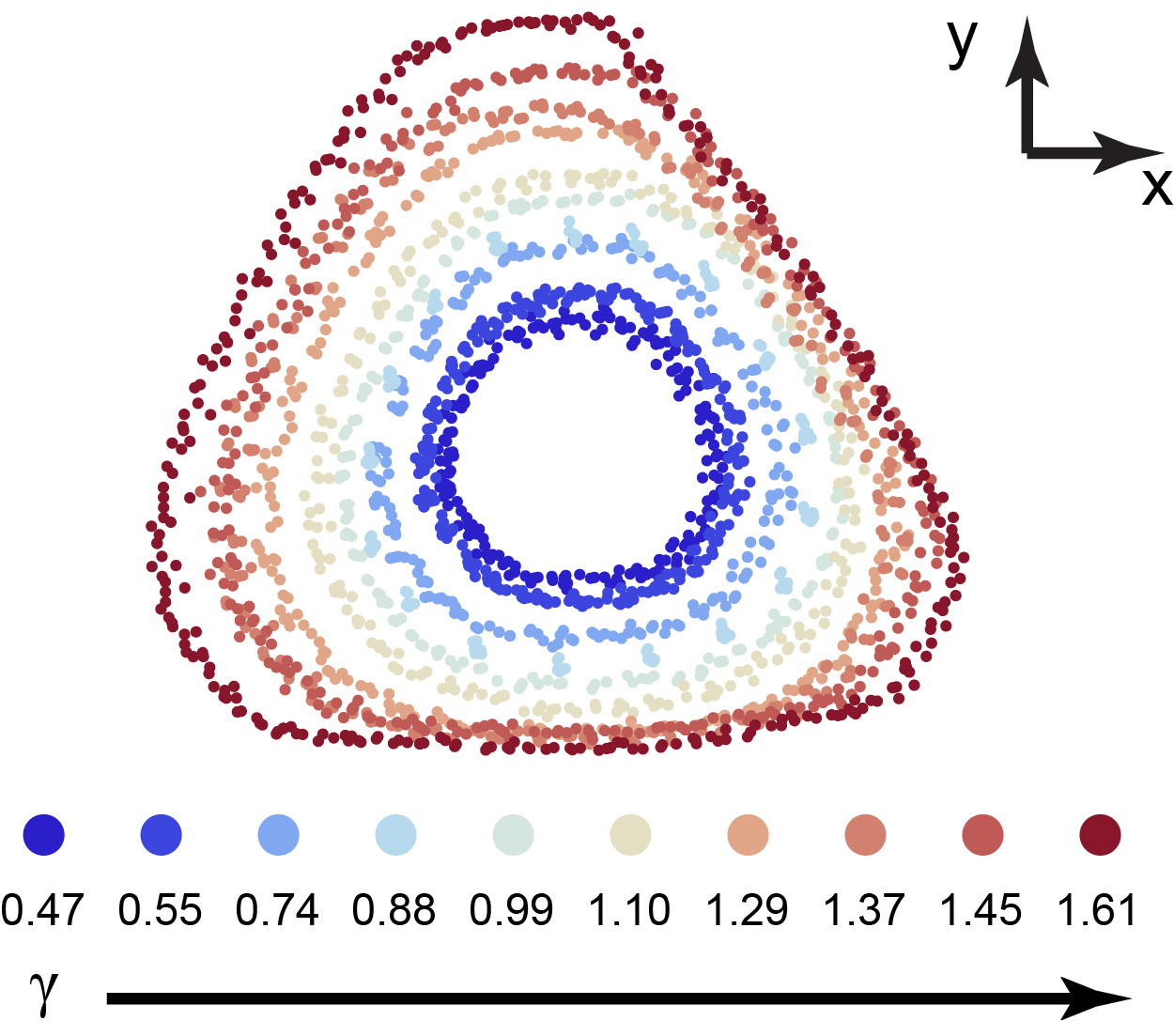}
\caption{\label{SI Fig5} 
2D projection of precessing gravity vector for different field amplitudes $\gamma$.
}
\end{figure*}
\newpage

\begin{figure*}
\includegraphics[width=\textwidth]{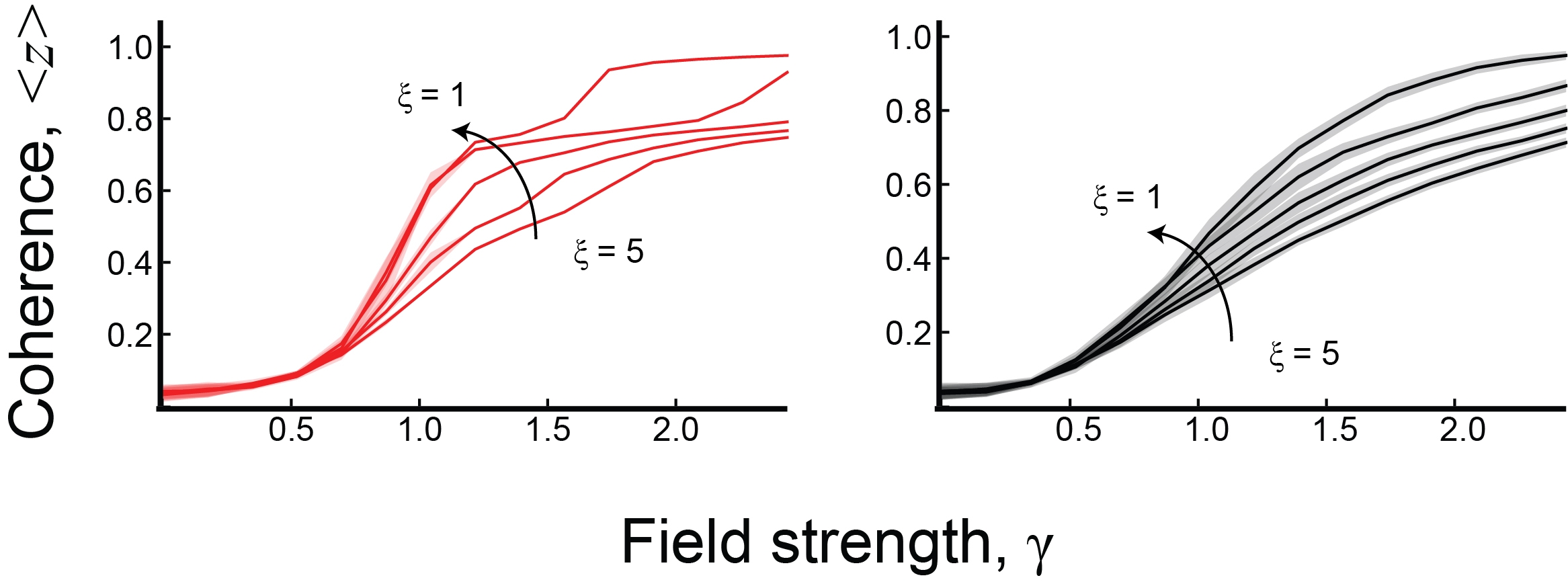}
\caption{\label{SI Fig6} 
Period averaged Kuramoto order parameter $\langle z \rangle$ as a function of field strength $\gamma$ for interaction strength ranging from $\xi = 1-5$. Systems with uniform droplet size (left) are compared against systems with 10\% disorder in droplet radii (right). Shaded regions represent standard deviation in $\langle z \rangle$ for $N=50$ samples.
}
\end{figure*}
\newpage

\begin{figure*}
\includegraphics[width=\textwidth]{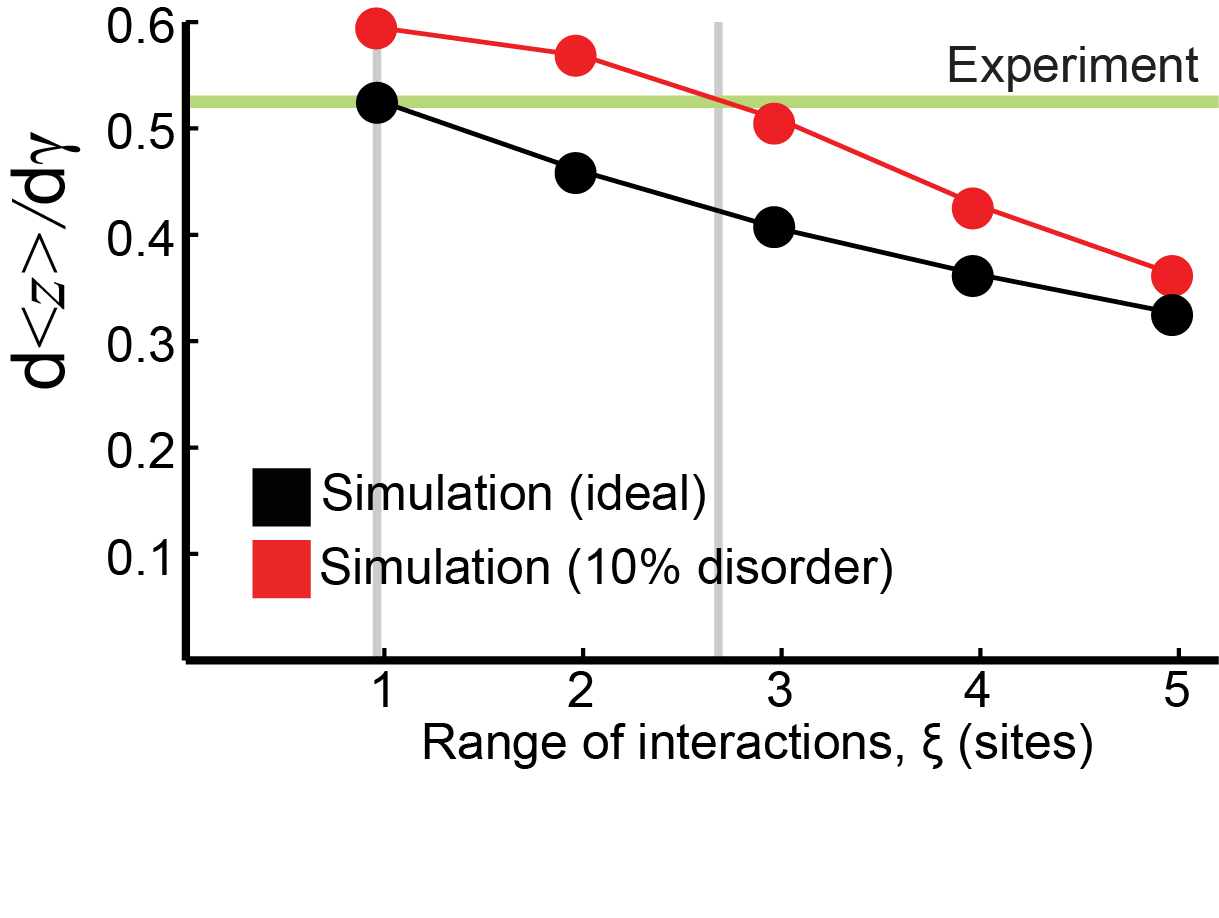}
\caption{\label{SI Fig7} 
Slope of Kuramoto order parameter in \textbf{SI Figure 6} plotted against number of interacting neighbors $\xi$. Horizontal line represents the equivalent calculation from experimental data shown in \textbf{Figure 3}. 
}
\end{figure*}
\newpage

\begin{figure*}
\includegraphics[width=\textwidth]{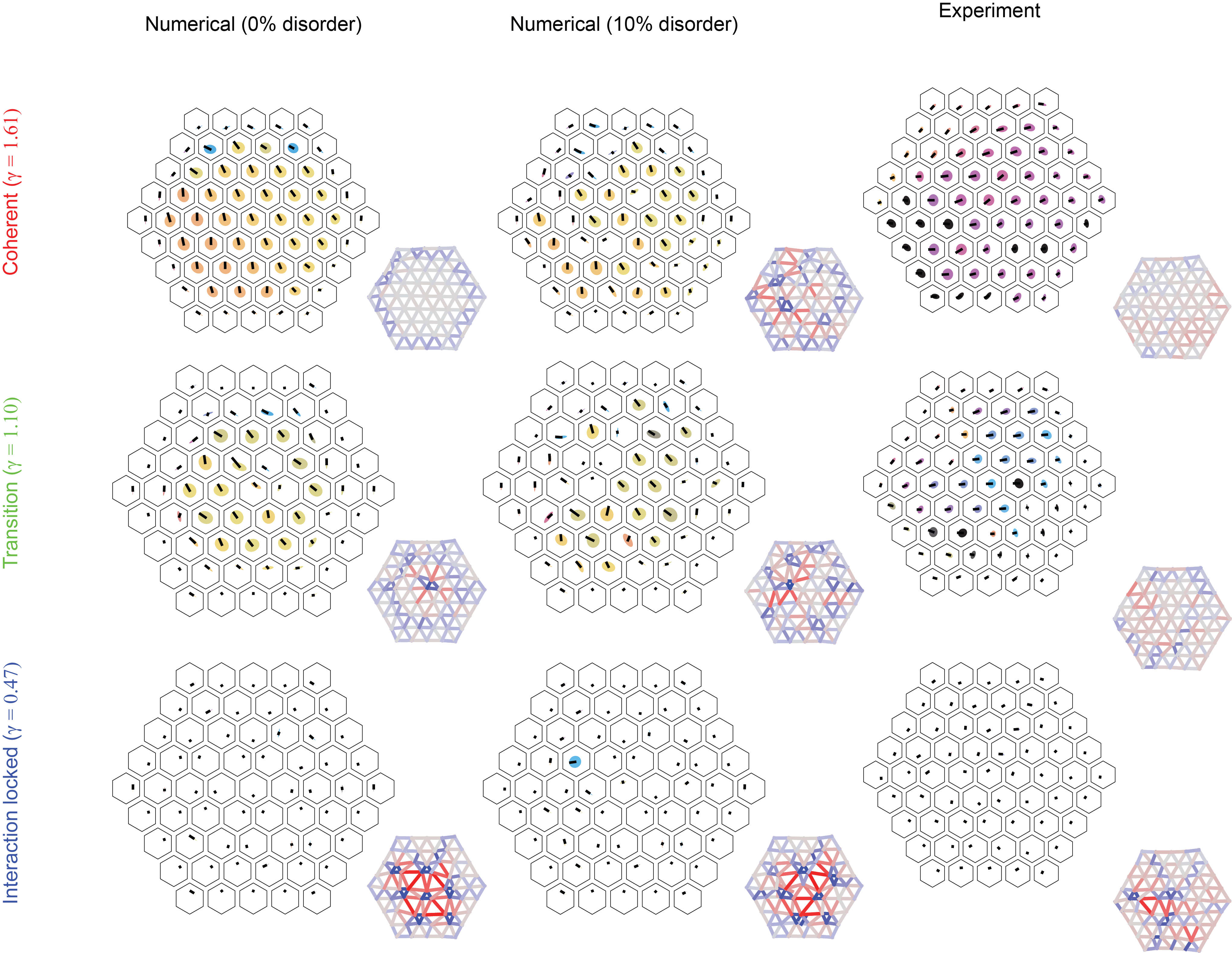}
\caption{\label{SI Fig8} 
Comparison of microscopic dynamics between ideal numerical simulations, numerical simulations with 10\% disorder in droplet radii, and experimental results. Large figures show droplet trajectories over one period with color indicating instantaneous phase of droplet. Insets show period averaged nearest-neighbor bond lengths (color scheme same as \textbf{Figure 3}).
}
\end{figure*}
\newpage

\begin{figure*}
\includegraphics[width=\textwidth]{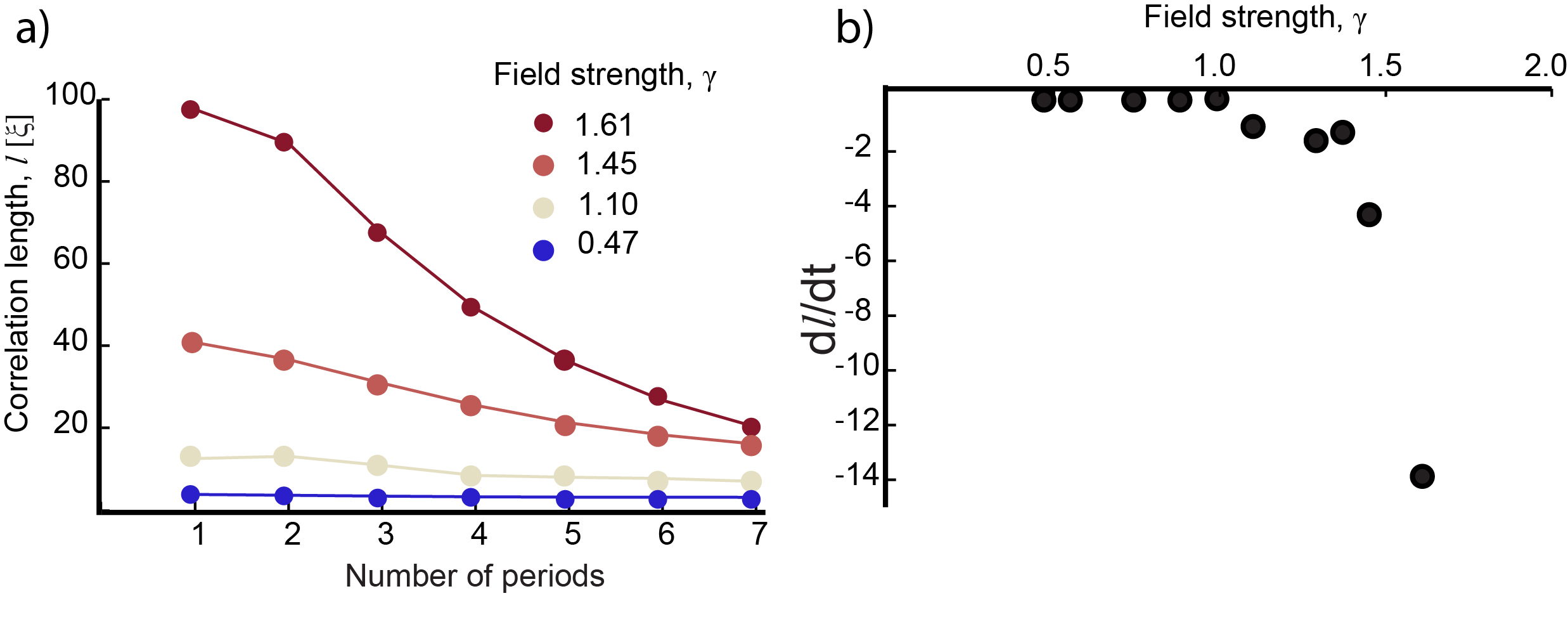}
\caption{\label{SI Fig9} 
\textbf{a}, Correlation length plotted as a function of number of periods shows that high correlation lengths will decay over time.
\textbf{b}, The rate of change in correlation length with time will diverge with increasing field strength $\gamma$
}
\end{figure*}
\newpage

\begin{figure*}
\includegraphics[width=\textwidth]{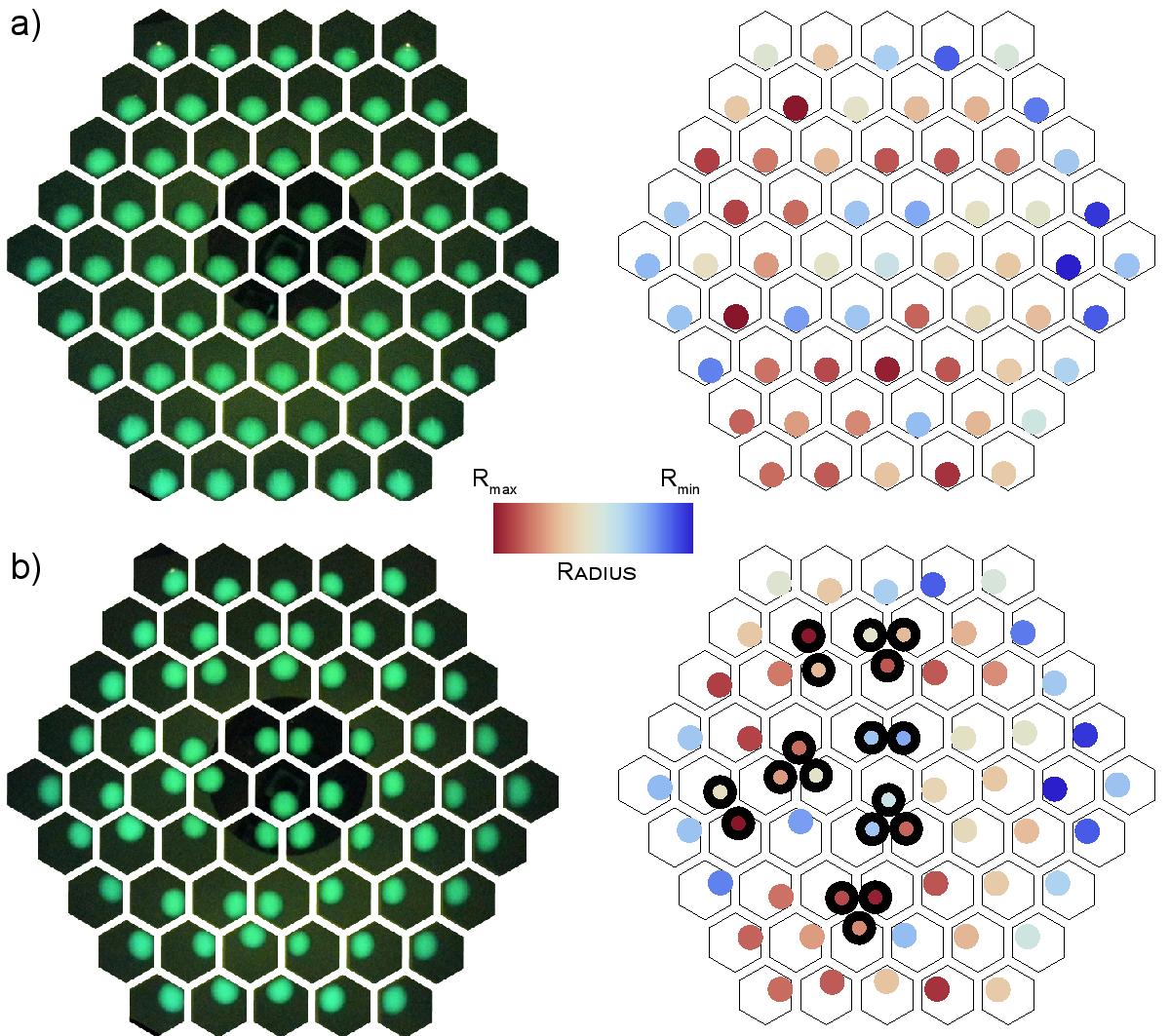}
\caption{\label{SI Fig10} 
\textbf{a} Initial and \textbf{b} final configurations for data shown in \textbf{Figure 4a} with raw data (left) compared against experiment-averaged radii (right). Vertex structures are indicated by black outlines. 
}
\end{figure*}
\newpage